\documentclass[12pt, draftclsnofoot, onecolumn]{IEEEtran}

\usepackage{cite}
\usepackage{graphicx}
\usepackage{amsmath, amsthm, amssymb}
\usepackage{algorithm, algorithmic}
\usepackage{color,xspace}
\usepackage{multirow, enumerate}
\usepackage{url}
\usepackage{tikz}
\usepackage{balance, verbatim}
\usetikzlibrary{arrows, shapes, positioning}

\theoremstyle{plain} 

\newtheorem{theorem}{Theorem}
\newtheorem{lemma}[theorem]{Lemma} 
\theoremstyle{definition} 

\newcommand{\setup}{\ensuremath{\mathsf{Setup}}}

\newcommand{\vct}[1]{\ensuremath{\mathbf{#1}}}

\newcommand{\mac}{\ensuremath{\mathsf{Mac}}}
\newcommand{\cbn}{\ensuremath{\mathsf{Combine}}}
\newcommand{\vrf}{\ensuremath{\mathsf{Verify}}}

\newcommand{\SMac}{\ensuremath{\mathsf{SpaceMac}}}
\newcommand{\HomMac}{\ensuremath{\mathsf{HomMac}}}

\newcommand{\HMac}{\ensuremath{\mathsf{HMAC}}}
\newcommand{\AES}{\ensuremath{\mathsf{AES}}}
\renewcommand{\triangleq}{\ensuremath{\overset{\text{def}}{=}}}

\newcommand{\iden}{\ensuremath{\mathsf{id}}}

\newcommand{\eff}[2]{\ensuremath{\mathbb{F}^{#1}_{#2}}}
\newcommand{\aug}[1]{\ensuremath{\mathsf{aug}(\vct{#1})}}

\newcommand{\lspan}[1]{\ensuremath{\mathsf{span}(\vct{#1})}}
\newcommand{\Prob}[1]{\ensuremath{\mathsf{Prob}[\vct{#1}]}}

\newcommand{\ie}{{\em i.e.}\xspace}
\newcommand{\ea}{{\em et al.}\xspace}
\newcommand{\eg}{{\em e.g.}\xspace}

\newcommand{\ncrypt}{\ensuremath{\mathsf{NCrypt}}\xspace}
\newcommand{\enc}{\ensuremath{\mathsf{Enc}}\xspace}
\newcommand{\dec}{\ensuremath{\mathsf{Dec}}\xspace}

\newcommand{\keygen}{\ensuremath{\mathsf{KeyGen}}\xspace}
\newcommand{\taggen}{\ensuremath{\mathsf{TagGen}}\xspace}
\newcommand{\genproof}{\ensuremath{\mathsf{GenProof}}\xspace}
\newcommand{\verifyproof}{\ensuremath{\mathsf{VerifyProof}}\xspace}
\newcommand{\chal}{\ensuremath{\mathsf{chal}}\xspace}

\newcommand{\ncaudit}{\ensuremath{\mathsf{NC\text{-}Audit}}\xspace}

\begin{document}

\title{Auditing for Distributed Storage Systems}

\author{
     Anh~Le,~\IEEEmembership{Member,~IEEE}
     Athina~Markopoulou,~\IEEEmembership{Senior Member,~IEEE}
        and~Alexandros~G.~Dimakis,~\IEEEmembership{Member,~IEEE}
\thanks{A. Le and A. Markopoulou are with UC Irvine. Emails: \{anh.le, athina\}@uci.edu}%
\thanks{A. G. Dimakis is with UT Austin. Email: dimakis@austin.utexas.edu}}

\maketitle

\begin{abstract}
Distributed storage codes have recently received a lot of attention in the community. Independently, another body of work has proposed integrity checking schemes for cloud storage, none of which, however, is customized for coding-based storage or can efficiently support repair. In this work,  we bridge the gap between these two currently disconnected bodies of work. We propose {\ncaudit}, a novel cryptography-based remote data integrity checking scheme,  designed specifically for network coding-based distributed storage systems. {\ncaudit} combines, for the first time, the following desired properties: (i) efficient checking of data integrity, (ii) efficient support for repairing failed nodes, and (iii) protection against information leakage when checking is performed by a third party. The key ingredient of the design of {\ncaudit} is a novel combination of {\SMac}, a homomorphic message authentication code (MAC) scheme for network coding, and {\ncrypt}, a novel chosen-plaintext attack (CPA) secure encryption scheme that preserves the correctness of {\SMac}. Our evaluation of {\ncaudit} based on a real Java implementation shows that the proposed scheme has significantly lower overhead compared to the state-of-the-art schemes for both auditing and repairing of failed nodes.
\end{abstract}

\begin{keywords}
Network Coding, Distributed Storage, Auditing, Integrity, Encryption, Security.
\end{keywords}

\section{Introduction}
\label{sec:audit-introduction}

\IEEEPARstart{T}{raditional} distributed storage architectures provide reliability through block replication, whose major disadvantage is the large storage overhead. As the amount of stored data is growing faster than hardware infrastructure, this becomes a major cost bottleneck. In contrast, coding techniques achieve higher data reliability with considerably smaller storage overhead~\cite{Ford2010}. For that  reason, coding techniques are under investigation for different distributed storage systems. Specifically, novel storage codes are currently being deployed in production cloud storage systems, such as Windows Azure~\cite{Huang2012}, analytics clusters (\eg, Facebook Analytics Hadoop clusters~\cite{Sathiamoorthy2013}), archival storage systems, and peer-to-peer storage systems like Cleversafe and Wuala \cite{cleversafe, wuala}.

Distributed storage codes operate by splitting files into blocks and creating additional parity blocks that provide fault tolerance. If the original file consists of $K$ blocks, an $(N, K)$ maximum distance separable (MDS) code is typically used to produce $N$ blocks to be stored individually on $N$ storage nodes, thus tolerating up to $(N-K)$ node failures. A well-known problem of classical erasure codes, like Reed-Solomon, is the so-called repair problem: when a single node fails, typically one block is lost from the file; however, the reconstruction of that single block requires reading and transferring $K$ blocks from other nodes. 

\begin{figure}[tp]
\centering
\begin{tikzpicture}
\node[scale=0.7] {
\begin{tikzpicture}

\tikzstyle{pre} =[
	<-,
	shorten <=1pt,
	>=stealth',
	semithick]

\tikzstyle{post}=[
	->,
	shorten >=1pt,
	>=stealth',
	semithick]

\tikzstyle{block} = [
	rectangle, 
      draw=black,
	fill=white, 
      thick,
	minimum width = 2.3cm]

\tikzstyle{circ} =[
	circle,
	draw=black,
	fill=black!10,
	thick,
	inner sep=0.2mm]

\tikzstyle{storage} = [
	rectangle, 
      draw=black, 
	fill=blue!10,
      thick,
	minimum width = 2.7cm,
	minimum height = 1.5cm,
	rounded corners]

\tikzstyle{fail} = [
	cross out, 
      draw=red, 
      line width=3pt,
	minimum width = 0.5cm,
	minimum height = 0.5cm]

\node[storage]	(n1)	at (0, 0)				{};
\path (n1)+(-2.5,0) node (n1name) 	{Node 1};
\path (n1)+(+0,+0.36) node (b1) [block] 		{$\vct{b}_1$};
\node[block, below=0.5mm of b1]       (b2)   	{$\vct{b}_2$};

\node[storage]	(n2)	at (0, -1.7)			{};
\path (n2)+(-2.5,0) node (n1name) 	{Node 2};
\path (n2)+(+0,+0.36) node (b3) [block] 		{$\vct{b}_3$};
\node[block, below=0.5mm of b3]       (b4)   	{$\vct{b}_4$};
\node[circ, right=6mm of n2]		(c)			{$+$};
\draw[post] (b3.east) to (c.west);
\draw[post] (b4.east) to (c.west);

\node[storage]	(n3)	at (0, -3.4)			{};
\path (n3)+(-2.5,0) node (n1name) 	{Node 3};
\path (n3)+(+0,+0.36) node (b5) [block] 		{$\vct{b}_1 + \vct{b}_3$};
\node[block, below=0.5mm of b5]       (b6)  	 	{$\vct{b}_2 + \vct{b}_4$};

\node[storage]	(n4)	at (0, -5.1)			{};
\path (n4)+(-2.5,0) node (n1name) 	{Node 4};
\path (n4)+(+0,+0.36) node (b7) [block] 		{$\vct{b}_2 + \vct{b}_3$};
\node[block, below=0.5mm of b7]       (b8)   	{$\vct{b}_1 + \vct{b}_2 + \vct{b}_4$};
\path (n4.west)+(+0.1,+0.3) node (x) [fail] {};

\node[block, right=17mm of b1]       	(b9)	{$\vct{b}_1$}
	edge [pre] (b1);
\node[block, right=5mm of c]			(b10)	{$\vct{b}_3 + \vct{b}_4$}
	edge [pre] (c);
\node[block, right=17mm of b6]       	(b11)	{$\vct{b}_2 + \vct{b}_4$}
	edge [pre] (b6);

\node[storage, right=4.5 of n4]		(n5)		{}
	edge [pre, bend right] (b9)
	edge [pre, bend right] (b10)
	edge [pre, bend right] (b11);
\path (n5)+(+0,+0.36) node (b12) [block] 	{$\vct{b}_2 + \vct{b}_3$};
\node[block, below=0.5mm of b12]       (b13)   	{$\vct{b}_1 + \vct{b}_2 + \vct{b}_4$};

\end{tikzpicture}
};
\end{tikzpicture}
\caption{Repairing a failed node \cite{Dimakis2011}: The original data consists of four blocks: $\vct{b}_1, \vct{b}_2, \vct{b}_3$ and $\vct{b}_4$. A $(4, 2)$ MDS code is used such that any 2 nodes can be used to restore the original data. Note that the repair involves combining blocks $\vct{b}_3$ and $\vct{b}_4$ and the repair bandwidth consists of 3 blocks instead of 4, where 4 is the amount of blocks needed to reconstruct the whole data.}
\label{fig:repair}
\vspace*{-15pt}
\end{figure}
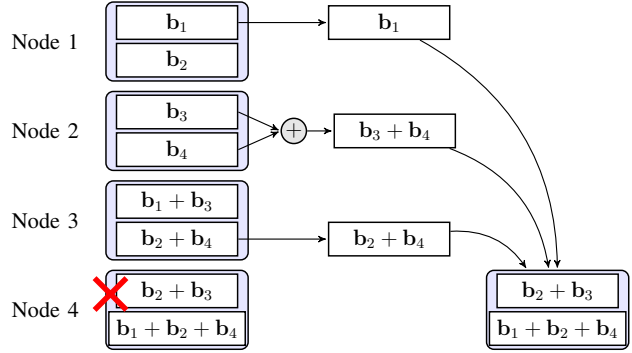

Novel storage codes that use network coding (NC) were recently developed to reduce this {\em repair bandwidth}. These distributed storage codes require significantly less than $K$ blocks to repair a single node failure and rely on network coding to perform in-network processing \cite{Dimakis2011, Dimakis2007}. Key ingredients of NC-based distributed storage codes include (i) storing {\em coded blocks}, \ie, linear combinations of original blocks that form the original data, and (ii) {\em block mixing} when repairing. An example is shown in Fig. \ref{fig:repair}. The repair bandwidth, however, is only one aspect of cloud storage.

Another practical aspect of cloud storage, besides the repair bandwidth, is data integrity checking. Integrity checking is extremely important for distributed storage systems, especially when data is stored with untrusted cloud providers. Data can be lost or corrupted for various reasons while users may remain completely unaware of for long periods of time. For example, storage errors, such as torn writes \cite{Krioukov2008} and latent errors \cite{Schroeder2010}, may damage data in a way that remains undetected. Cloud storage providers may also have incentives to misbehave, \eg,  misreport data loss incidents in order to maintain their reputation \cite{Ateniese2007, Shah2007, Wang2009}. This problem is further exacerbated in systems that use coding because corrupted data can propagate to multiple nodes during repair re-encoding \cite{Chen2010}. Therefore, it is important for the user  to be able to audit the integrity of the data stored on the cloud. 

Another complication is that frequent integrity checking of large data sets may be out of the ability or budget of users with limited resources \cite{Wang2009, CloudSecurity}. As a result, users often resort to a third party to perform audits on their behalf \cite{Ateniese2007, Wang2009, Shacham2008, Juels2007}. In this latter case, it is important that the auditing protocols are privacy-preserving, \ie, do not leak information to the third party \cite{Wang2009, CWang2010}. Indeed, users can leverage data encryption to protect their data before outsourcing it \cite{Juels2007}. However, data encryption should be complementary and orthogonal to integrity checking protocols. In other words, the auditing protocol should not introduce new vulnerabilities of unauthorized data leakage. Furthermore, the users may want to outsource unencrypted instead of encrypted data to support more efficient and complex computations.  

As a result, auditing for distributed systems that use modern NC-based storage codes is an important emerging problem. Despite the rich literature on auditing protocols for general distributed and cloud storage~\cite{Juels2007, Ateniese2007, Shah2007, Ateniese2008, Shacham2008, Bowers2009, Bowers2009Hail, Erway2009, Wang2009, QWang2009, CWang2010, Yu2010, Wang2011}, there have been very few auditing protocols  for NC-based distributed storage systems \cite{Chen2010, Dikaliotis2010}. These protocols, however, are generic in the sense that they do not specifically exploit coding properties  for efficient integrity checking \cite{Chen2010}. Moreover, they do not prevent data leakage \cite{Chen2010, Dikaliotis2010}. Most importantly,  they do not efficiently support repair, which is the main advantage of NC-based storage systems when compared to other storage systems.

In this work, we propose a symmetric key-based cryptographic protocol, called {\ncaudit}, to check for the integrity of data stored on an NC-based distributed storage system. To the best of our knowledge, this is the first scheme proposed for NC-based systems that possesses all the following desired properties:
\begin{enumerate}[(i)]
\item {\bf Efficient Integrity Checking}: The integrity check incurs a small bandwidth and computational overhead (on the order of milliseconds). It guarantees that, with high probability, the storage provider passes the integrity check if and only if it possesses the data. The proposed protocol also supports unlimited number of checks.
\item {\bf Efficient Support for Repair}: The repair of failed nodes require negligible bandwidth (no data download) as well as computation for maintaining the metadata used by the integrity checking.
\item {\bf Efficient Privacy Protection}: A third party auditor cannot learn any information about the user data through the checking protocol (except for the metadata used by the integrity checking). This privacy preserving property incurs a small bandwidth ($<$ 1\%) and computational overhead (on the order of milliseconds).
\end{enumerate}

We would like to emphasize that, independently of (iii), properties (i) and (ii) together are already useful to users who could and prefer to audit the data themselves. {\ncaudit} is the first protocol that possesses (i) and (ii) at the same time. {\ncaudit} achieves these properties by fully exploiting network coding in its design. The main novelty of {\ncaudit} come from a careful combination of  {\SMac}  -- a homomorphic message authentication code (MAC) that was previously specifically designed for network coding \cite{LeLocate2010, LeInter2012}, and  {\ncrypt} -- a novel chosen-plaintext attack (CPA) secure encryption scheme that we  custom designed, in this work, to operate in synergy with and preserve the correctness of {\SMac}. 

We implemented {\ncaudit} in Java, utilizing our previous implementation of {\SMac} \cite{LeInter2012}. Our evaluation of {\ncaudit} shows that it has very low computational overhead. In particular, when performing an audit, both the storage node and the auditor only need to spend a few milliseconds. Furthermore, the auditor's overhead is much less than that of the state-of-the-art approach for NC-based storage systems \cite{Chen2010}, which is on the order of seconds.

The rest of the paper is organized as follows. In Section \ref{sec:audit-related_work}, we discuss related work. In Section \ref{sec:audit-formulation}, we formulate the problem and describe the threat model. In Section \ref{sec:audit-auditing}, we describe the auditing framework and the key building blocks of {\ncaudit}, namely {\SMac} and {\ncrypt}, before presenting {\ncaudit} itself. In Section \ref{sec:audit-repair}, we show how {\ncaudit} efficiently supports repair. In Section \ref{sec:audit-security}, we analyze the security of {\ncaudit}. In Section \ref{sec:audit-evaluation}, we evaluate its storage, bandwidth, and computational efficiency. In Section \ref{sec:audit-conclusion}, we conclude the paper.

\section{Related Work}
\label{sec:audit-related_work}

\subsection{Integrity Checking for Remote Data}
\label{subsec:related-checking}
There has been a rich body of work on integrity checking for remote data \cite{Juels2007, Ateniese2007, Shah2007, Ateniese2008, Shacham2008, Bowers2009, Erway2009, Wang2009, QWang2009, CWang2010, Yu2010, Wang2011}, commonly known as {\em Proof of Retrievability} and {\em Proof of Data Possession}. 

{\flushleft \bf Proof of Retrievability (POR).}\quad In \cite{Juels2007}, Juels and Kaliski introduced the notion of POR, where a POR enables a client (verifier) to determine that the server (prover) possesses a file or data object. Furthermore, a successful execution of POR would allow a verifier to extract the file from the proof. The main POR scheme presented there uses {\em sentinels}, \ie, small check blocks, that are inserted into the outsourced data to guard against large file corruption. At the same time, it also utilizes error correcting codes to protect against small file corruption. This scheme can only handle a limited number of queries, which has to be fixed a priori. In contrast, {\ncaudit} does not use sentinels and supports unlimited number of queries.

In \cite{Shacham2008}, Shacham and Waters proposed two POR schemes with full proofs of security and extract-ability. The first one, built on Boneh-Lynn-Shacham (BLS) signatures, provides public verifiability. The second one, built on pseudorandom functions (PRFs), provides private verifiability. Recently, Bowers \ea \cite{Bowers2009Hail} proposed HAIL,  an improvement of existing POR schemes that allows for performing data integrity checking with multiple servers against stronger, mobile adversaries. 

These schemes  \cite{Shacham2008, Bowers2009Hail} exploit homomorphic properties to aggregate authenticator values to improve the audit efficiency. {\ncaudit} also exploits homomorphic properties (of \SMac) and provides private verifiability. In terms of extract-ability, {\ncaudit} is different from existing approaches, \eg, \cite{Shacham2008}, in that {\ncaudit} exploits the inherent embedded coding coefficients in the stored blocks to perform the extraction. Meanwhile, \cite{Shacham2008} relies on additional erasure codes (pre-applied to the data) for the extraction.

{\flushleft \bf Proof of Data Possession (PDP).}\quad The notion of PDP was introduced by Ateniese \ea \cite{Ateniese2007}. The PDP scheme in \cite{Ateniese2007} uses homomorphic RSA signatures to generate verification tags. The data possession guarantee provided by this scheme is under the RSA and KEA1 assumptions in the random oracle model. Earlier in \cite{Schwarz2006}, Schwarz and Miller proposed using a combination of both erasure-correcting coding and algebraic signatures (homomorphic hashes) to perform integrity checking for remote data. As discussed in \cite{Shacham2008}, the notion of PDP is considered to be weaker than POR. This is because in POR, a successful audit guarantees that all the data can be extracted while in PDP, only a certain percentage of the data (\eg, 90\%) is guaranteed to be available. Integrity checking for groups with efficient user revocation was recently introduced in \cite{Wang2013}. We will show that {\ncaudit} provides the stronger data possession checking with data extraction as in POR (Section \ref{audit-subsec:possession}).

{\flushleft \bf Data Modification.}\quad In \cite{Ateniese2008}, Ateniese \ea proposed a symmetric-key based checking scheme that supports data modification. This scheme is built on regular PRFs, hash functions, and encryptions. It provides private verifiability and supports a limited number of queries. In \cite{Erway2009}, Erway \ea proposed an auditing scheme built on rank-based authenticated skip lists and requires the storage server to maintain the lists for verification. In \cite{QWang2009}, Wang \ea~proposed a public auditing scheme that uses a combination of the BLS-based scheme in \cite{Shacham2008} and Merkle Hash Tree (MHT). 

In practice, most current deployments of distributed storage codes \cite{Sathiamoorthy2013, Huang2012} initially set all files to replication mode. When certain files become \textit{cold} (\ie, rarely accessed and modified) the replicated blocks are deleted and corresponding parity blocks are created. This dynamic switching of files from replication to coding allows distributed storage systems to benefit from the high performance of replication for \textit{hot} files and the storage benefits of coding for \textit{cold} files. Interestingly, in most analytics clusters and cloud storage systems, the vast majority of data seem to be \textit{cold} \cite{Sathiamoorthy2013, Huang2012}. Therefore, we do not expect data modification to be a critical operation for encoded data. {\ncaudit} provides some preliminary support for data modification, and the details can be found in the Appendix.




{\flushleft \bf Privacy Preserving.}\quad In \cite{Shah2007}, Shah \ea~proposed an auditing protocol that is privacy preserving. This protocol first encrypts the data and then sends a number of message authentication code (MAC) tags of the encrypted data to the auditor. The auditor verifies both the outsourced data and the outsourced encryption key. This approach only works on encrypted files. It also requires the auditor to maintain states and supports only limited number of audits. In \cite{CWang2010}, Wang \ea~ proposed a privacy preserving auditing protocol that has public verifiability. This protocol can be considered an extension of the BLS-based protocol in \cite{Shacham2008}. In this approach, the aggregated (proving) block sent by the storage server is masked with a random element to protect the privacy of the block. {\ncaudit} is explicitly designed to provide privacy preserving-auditing (Section \ref{audit-subsec:audit-auditing} and \ref{audit-subsec:privacy}). Different from \cite{CWang2010}, {\ncaudit} relies on symmetric-key cryptographic primitives instead of public-key ones, and thus it provides private instead of public auditing.

Finally, we stress that none of the schemes described above was customized for NC-based storage. In particular, they do not provide efficient support for node repair. {\ncaudit} was designed to achieve all the above properties: providing proof of retrievability and privacy-preserving auditing while efficiently supporting node repair.

\subsection{Integrity Checking for NC-based Storage Systems}
\label{audit-subsec:nc_based_check}

{\flushleft \bf NC-based Storage Systems.}\quad The benefits of network coding for distributed storage were first formalized by the work of Dimakis \ea \cite{Dimakis2007}. In particular, in \cite{Dimakis2007}, the authors proposed the notion of {\em regenerating codes} and show that they can significantly reduce the repair bandwidth. This work showed the fundamental tradeoff between node storage and repair bandwidth and proposed regenerating codes that can achieve any point on the optimal tradeoff curve. A survey on recent advances in NC-based storage system can be found at \cite{Dimakis2011}. A wiki on NC-based storage cloud is maintained at \cite{DimakisWiki}. {\ncaudit} is designed to {\em fully support regenerating codes}.

An NC-based distributed file system (NCFS) is proposed in \cite{Hu2011}. One of the first implementations of NC-based storage cloud is NCCloud by Hu \ea \cite{Hu2012}. In particular, NCCloud is a proxy-based system for multiple-cloud storage. It utilizes a functional minimum-storage regenerating code to provide cost-effective repair for a permanent single-cloud failure. This efficient repair is achieved without the cost of storage or redundancy level. NCCloud prototype was deployed on top of Windows Azure Storage.

{\flushleft \bf Integrity Checking Schemes for NC-Based Storage Systems.}\quad There have been only a few number of work that provide remote data checking for NC-based storage. In \cite{Dikaliotis2010}, Dikialotis \ea~proposed an integrity checking scheme that utilizes the error-correction capabilities of the storage system. This scheme aims to detect errors with a very small amount of bandwidth. The key technique for reducing the bandwidth is to project data blocks onto a small random vector. This checking scheme is inherently different from {\ncaudit} as it relies on the communication between the auditor and multiple nodes to perform a single check while {\ncaudit} does not. Moreover, this scheme is information-theory based while {\ncaudit} leverages cryptographic primitives to provide the checking.

A more recent integrity checking scheme for NC-based storage was proposed in \cite{Chen2010}. In this work, Chen \ea~adopted the symmetric-key based scheme that Shacham and Waters proposed for regular cloud storage \cite{Shacham2008} with minor modification. In particular, based on the symmetric-key based scheme in \cite{Shacham2008}, the scheme in \cite{Chen2010} proposed to encrypt the coding coefficients of the outsourced encoded blocks to prevent {\em replay attacks}, where a malicious storage node may store old (incorrect) encoded blocks instead of the new (correct) encoded blocks as required by the repair \cite{Chen2010}. {\ncaudit} overcomes this attack by requiring the user/auditor to store the coding coefficients, which is also needed for the repair process and only occupies a negligible amount of storage (see Section \ref{audit-subsec:storageOverhead}). 

What really sets {\ncaudit} apart from \cite{Chen2010} is that {\ncaudit} fully exploits network coding for integrity checking. In particular, the scheme proposed in \cite{Chen2010} relies on two independent logical representation of file blocks for two different purposes: data possession checking and network coding operation. Because of this, during the repair process, the user has to download blocks from the remaining healthy nodes to compute the integrity checking data for the new coded blocks (to be stored at the recovery node). This approach puts heavy bandwidth and computational overhead on the user. In contrast, {\ncaudit} uses a {\em single} representation for both purposes and thereby achieving integrity checking while eliminating the heavy user's bandwidth and computational overhead. Details of how {\ncaudit} support efficient repair are provided in Section \ref{audit-subsec:repair}. Furthermore, the scheme in \cite{Chen2010} does not support privacy-preserving auditing while {\ncaudit} does. We provide detailed performance comparison between {\ncaudit} and \cite{Chen2010} in Section \ref{sec:audit-evaluation}.

Finally, a recent work by Cao \ea \cite{Cao2012} proposed an LT codes-based storage system with an integrity checking and an exact repair schemes; however, it neither supports functional repair \cite{Dimakis2007} (discussed in Section \ref{audit-subsec:repair}) nor privacy-preserving auditing.

{\flushleft \bf Other Security Issues.}\quad 
Other security problems for NC-based storage include protecting the privacy and integrity of the blocks while repairing. The work in \cite{Pawar2010} and \cite{Rouayheb2010} prevents eavesdroppers from accessing/decoding all the data. In \cite{Pawar2010}, Pawar \ea~provide an explicit code construction that achieves the secrecy capacity for the bandwidth-limited regime of the storage systems under repair dynamics. \cite{Rouayheb2010} analyzes the effects of interaction between the storage nodes on the amount of data revealed to the eavesdroppers. The work in \cite{Pawar2011a} provides upper bounds on the maximum amount of information that can be stored safely when there are malicious nodes.

In \cite{Pawar2011a} and \cite{Buttyan2010}, the authors provide protection against pollution attacks during the repair.  In \cite{Buttyan2010}, Buttyan \ea~provide a lightweight, pollution-resilient decoding algorithm that is capable of finding adversarial blocks. The scheme in \cite{Chen2010} also protects the repair phase against pollution attacks, \ie, preventing remaining nodes from sending corrupted data to the new (recovering) node. Dealing with pollution attacks is out of the scope of this work. We refer the reader to the rich literature, including our previous work, that deal with pollution attacks \cite{Gkantsidis2006, Agrawal2009, Li2010, Boneh2009, Zhang2011, LeTESLA2011, LeJSAC2011}. 

\subsection{This Work in Perspective}
A preliminary version of this work has appeared in NetCod 2012 \cite{Le2012}. In this paper, we provide the following revisions and extensions of the previous version: We revise and provide complete proofs of all lemmas and theorems; we described in detail a repair process; we discuss and compare our storage overhead to prior work \cite{QWang2009, CWang2010, Chen2010}; finally, we provide a comprehensive discussion of related literature.

\section{Problem Formulation}
\label{sec:audit-formulation}

\subsection{System Model and Operations}
\label{audit-subsec:system_model}

\begin{figure}[t]
\centering
\includegraphics[width=7.5cm]{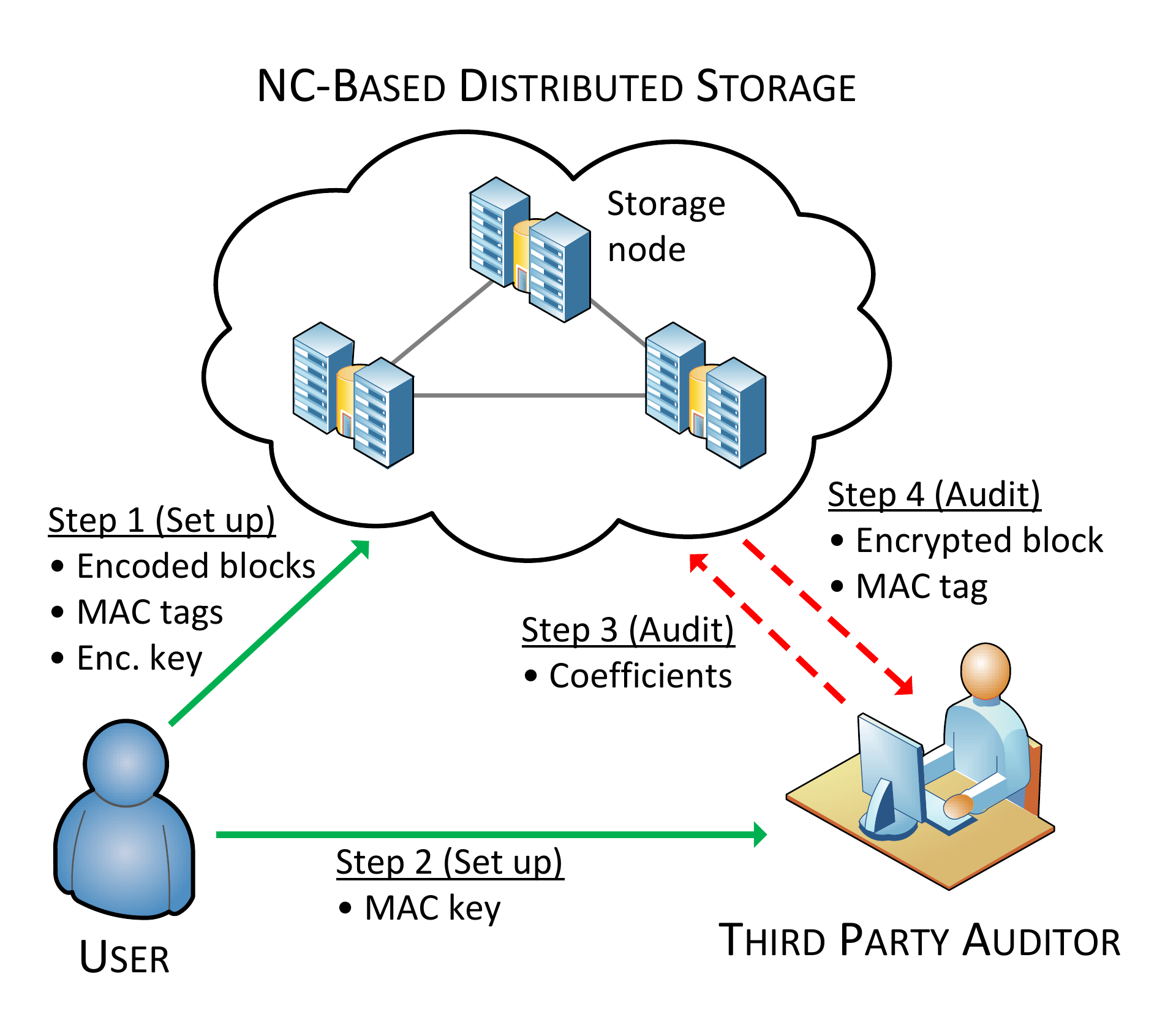}
\vspace{-10pt}
\caption{Parties and Steps Involved in {\ncaudit}.}
\label{fig:CSP}
\vspace{-15pt}
\end{figure}

Fig. \ref{fig:CSP} illustrates an overview of {\ncaudit}. We consider a cloud storage service that involves three entities: a user, NC-based storage nodes, which make up the storage cloud, and a third party auditor (TPA). The user distributes his/her data on the storage nodes. The user resorts to a TPA to check for the integrity of the data stored at each node; at the same time, he/she does not want the TPA to learn about the outsourced data. We assume that {\em the user is responsible for repairing of a failed node.} The user here acts as a proxy that manages the storage nodes as in the case of NCCloud  \cite{Hu2012}. Our work is also applicable to scenarios where there is a cloud service provider, who is independent from the user and acts as the proxy.

The user follows the following basic steps to store his/her data on the storage cloud. We adopt the notation used in \cite{LeTESLA2011}. Denote the original file by $\mathcal{F}$. The user first divides $\mathcal{F}$ into $m$ blocks, $\hat{\vct{b}}_1, \cdots, \hat{\vct{b}}_m$. Each block is a vector in an $n$-dimensional linear space $\eff{n}{q}$, where $\eff{}{}$ is a finite field of size $q$. To facilitate the decoding, the user then augments each block $\hat{\vct{b}}_i$ with its $m$ {\em global coding coefficients}. The resulting blocks, $\vct{b}_i$, have the following form:
\[
\vct{b}_i = (\, \overbrace{\textrm{---}\vct{\hat{b}}_i\textrm{---}}^n, \overbrace{\underbrace{0, \cdots, 0, 1}_i, 0, \cdots, 0}^m)\,\in \mathbb{F}^{n+m}_q\,.
\]
We call $\vct{b}_i$ {\em source blocks} and the space spanned by them {\em source space}, denoted by $\Pi$. We use \aug{\vct{b}_i} to denote the coefficients of $\vct{b}_i$. Typically, $n \gg m$, and this presentation is also called an $n$-extended version of a storage code \cite{Dikaliotis2010}. 

The user then creates a number of encoded blocks using an appropriate linear coding scheme for the desired reliability, \eg, an array MDS Evenodd code is used in Fig. \ref{fig:repair}. Each encoded block is a linear combination of the source blocks. Note that if an encoded block $\vct{e}$  equals $\sum_{i=1}^m \alpha_i \, \vct{b}_i$, then the last $m$ coordinates of $\vct{e}$ are exactly the coding coefficients $\alpha_i$'s. These encoded blocks are then distributed across the $N$ storage nodes of the storage cloud. Let $M$ be the number of encoded blocks stored at a storage node, $P$ be the number of healthy nodes that need to send the (encoded) repair blocks, and $Q$ be the number of repair blocks each healthy node needs to send to the new node. In the example given in Fig. \ref{fig:repair}, $m=4$, $N=4$, $M=2$, $P=3$, and $Q=1$.

\subsection{Threat Model}
\label{audit-subsec:threat_model}

We adopt the threat model considered in \cite{CWang2010} and \cite{Yu2010}. In particular, we consider semi-trusted storage nodes that behave properly and do not deviate from the prescribed auditing protocol. However, for their own benefits, they may deliberately delete rarely accessed, archival user's data to reduce operational cost; they may also decide to hide data corruptions, caused by either internal or external factors to maintain reputation. For clarity, we focus our discussion on a single storage node except when discussing the repair process.

We assume that the TPA, who is in the business of auditing, is reliable and independent. We assume that the TPA does {\em not collude} with the storage node during the auditing process to hide data corruption. This is a standard assumption when relying on a TPA for integrity checking \cite{QWang2009, CWang2010, Wang2013}. The TPA, however, must not be able to learn any information about the user's data through the auditing process, aside from the metadata needed for the auditing, as in \cite{CWang2010}. In order words, the auditing protocol should not introduce a data leakage vulnerability. Similar to standard applications of cryptographic protocols, we assume that both the node and the TPA are fully aware of all the cryptographic constructions and protocols used; however, their runtime is polynomial in the security parameter.

\section{Auditing Scheme}
\label{sec:audit-auditing}

\subsection{Definitions and Auditing Framework}
\label{audit-subsec:frame_work}

We follow the literature of integrity checking of remote data \cite{Ateniese2007, Juels2007, Shacham2008, Bowers2009, CWang2010} and adapt the proposed framework to our privacy-preserving auditing system. In particular, we consider an auditing scheme which consists of four algorithms: 
\begin{itemize}
\item $\keygen(1^\lambda) \rightarrow (k_v, k_e)$\quad is a key generation algorithm that is run by the user to setup the scheme. It takes a security parameter $\lambda$ as input and outputs two different private keys: $k_v$ used to generate {\em verification} metadata, and $k_e$ used to {\em encrypt} the possession proof.

\item $\taggen(\vct{e}, k_v) \rightarrow t$\quad is an algorithm run by the user to generate the verification metadata. It takes as input a coded block, $\vct{e}$, a private key, $k_v$, and outputs a verification tag of $\vct{e}$, $t$.

\item $\genproof(k_e, (\vct{e}_1, \cdots, \vct{e}_M), (t_{\vct{e}_1}, \cdots, t_{\vct{e}_M}), \chal) \rightarrow V$\quad is run by the storage node to generate a proof of possession. It takes as input a private key, $k_e$; coded blocks stored at the node, $\vct{e}_1, \cdots, \vct{e}_M$; their corresponding verification metadata, $t_{\vct{e}_1}, \cdots, t_{\vct{e}_M}$; and a challenge, \chal, which includes block indices and coding coefficients. It outputs a proof of possession, $V$, for the coded blocks  determined by \chal.

\item $\verifyproof(k_v, \chal, V) \rightarrow \{1, 0\}$\quad is run by the TPA in order to validate a proof of possession. It takes as inputs a private key, $k_v$, a challenge, $\chal$, and a proof of possession $V$. It returns 1 (success) if $V$ is the correct proof of possession for the blocks determined by $\chal$ and 0 (failure) otherwise.
\end{itemize}

An auditing system can be constructed from the above algorithms and consists of two phases:
\begin{itemize}
\item {\em Setup}: The user initializes the security parameters of the system by running \keygen. The encoded blocks are prepared as previously described in Section \ref{audit-subsec:system_model}. The user then runs {\taggen} to generate verification metadata for each encoded block. Afterwards, both the encoded blocks and verification metadata are uploaded to the storage node. The encoded blocks are then deleted from the user's local storage. Finally, the user sends metadata needed to perform the audit to the TPA.
\item {\em Audit}: The TPA issues an audit message, \ie, a {\chal}, to the storage node to make sure that the node correctly stores its assigned coded blocks. The node generates a proof of possession for the blocks specified in {\chal} by running {\genproof}, and it sends the possession proof back to the TPA. Finally, the TPA runs {\verifyproof} to verify the possession proof it receives.
\end{itemize}

\subsection{Basic Scheme and Key Techniques}
\label{audit-subsec:basic_scheme}

Here we describe the most basic scheme that supports remote data checking and show that it does not provide the desired properties. This basic scheme is also described in \cite{Ateniese2007}. Afterwards, we describe how we improve this basic scheme to arrive at our proposed scheme.

{\flushleft \bf The Basic Scheme.} During the {\em Setup} phase, the user precomputes a Message Authentication Code (MAC) tag, $t_i$, for each coded block, $\vct{e}_i$, using a secret key, $k_v$, and a standard MAC scheme, \eg, \HMac. The user then uploads both the tags and the coded blocks to the storage node and sends $k_v$ to the TPA. During the {\em Audit} phase, to verify that the node stores $\vct{e}_i$ correctly, the TPA issues a request for $\vct{e}_i$. The node then sends $\vct{e}_i$ and its tag $t_i$ to the TPA. The TPA can use $k_v$ and $t_i$ to check for the integrity of $\vct{e}_i$.
Although providing the possession checking, this scheme suffers from many drawbacks: 
\begin{itemize}
\item It is inefficient in both computation and communication since the computation and bandwidth overhead increases linearly in the number of checked blocks. 
\item It does not efficiently support {\em node repair} \cite{Dimakis2011, Dimakis2007}: It requires the user to download all the blocks necessary to compute the new (recovering) blocks. The user then computes verification tags for all the new blocks, essentially re-setting up the storage node.
\item It violates privacy because the TPA learns about the blocks. A straightforward way to provide privacy is to encrypt the response block using a standard encryption scheme, \eg, \AES. However, in this case, the TPA will not be able to verify the integrity of the original block because the provided tag is not computed on the encrypted block but on the original block.
\end{itemize}

{\flushleft \bf Key Techniques.} We improve the basic scheme to arrive at our proposed scheme by leveraging a novel combination of (i) a homomorphic MAC scheme and (ii) a novel encryption scheme that exploits properties of linear network coding. 

In detail, we adopt \SMac, a homomorphic MAC scheme that we previously designed specifically for network coding \cite{LeLocate2010, LeJSAC2011}. We use {\SMac} to generate verification tags. With \SMac, the integrity of multiple blocks can be verified with the computation and communication cost of a single block verification, thanks to the ability to combine blocks and tags. {\SMac} also facilitates repair as verification metadata at a newly constructed node can be computed efficiently from existing metadata at healthy nodes.

We custom design a novel encryption scheme, called \ncrypt, to protect the privacy of the response blocks. {\ncrypt} is constructed in a way  that preserves the correctness of {\SMac}: A response block, even when encrypted, can be used by the TPA for the integrity check. We stress that it is not possible to use other standard encryption schemes, such as {\AES}, in place of \ncrypt, because they will break the {\SMac} integrity verification. The reason is that in general, a MAC tag computed on a data block can only be used to verify the integrity of the block upon the reception of the tag and the data block, but it cannot be used when the encrypted data block is received instead of the original block. 

Formally, let $(\enc, \dec)$ denote a symmetric-key encryption scheme and $(\mac, \vrf)$ denote a MAC scheme. Let $\vct{e}$ be an (encoded) data block, and $k_e$ and $k_v$ be the keys for the encryption and MAC schemes. Let $\vct{c} = \enc(k_e, \vct{e})$ and $t = \mac(k_v, \vct{e})$. The encryption and MAC schemes are compatible with each other when $\vrf(k_v, \vct{c}, t)$ outputs 1 if and only if $\vct{c} = \enc(k_e, \vct{e})$ and outputs 0 otherwise.

The main novelty of {\ncrypt} lies in its compatibility with \SMac: It is carefully designed to maintains both the correctness of {\SMac} (Theorem \ref{thm:correctness}) as well as the security of {\SMac} (Theorem \ref{thm:PossessionSpaceMac}). {\ncrypt} employs the random linear combination technique of network coding and is semantically secure under a chosen-plaintext attack (CPA-secure). Next, we describe how we use {\SMac} and {\ncrypt} in detail.

\subsection{The Homomorphic MAC: \SMac}
\label{audit-subsec:spacemac}

In prior work, we originally designed $\SMac$ and used it to combat pollution attacks in network coding \cite{LeLocate2010, LeJSAC2011, LeTESLA2011, LeInter2012}. {\SMac} was inspired by and an improvement of another homomorphic MAC scheme, \HomMac, proposed by Agrawal and Boneh \cite{Agrawal2009}. The novelty of {\SMac} and a detailed comparison between the two schemes can be found in \cite{LeLocate2010, LeJSAC2011}. Here, we adopt $\SMac$ to support the aggregation of file blocks and tags to allow for efficient auditing (similar to \cite{Shacham2008, Bowers2009Hail}). Furthermore, as we show in Section \ref{sec:audit-repair}, {\SMac} also facilitates efficient node repairs.

{\flushleft \bf Definition.} A ($q, n, m$) homomorphic MAC scheme is defined by three probabilistic, polynomial-time algorithms: $\mac$, $\cbn$, and $\vrf$. The $\mac$ algorithm generates a tag for a given block; the $\cbn$ algorithm computes a tag for a linear combination of some given blocks; and the $\vrf$ algorithm verifies whether a tag is a valid tag of a given block.
\begin{itemize}
\item $\mac(k, \text{id}, \mathbf{e})$:
  \begin{itemize}
  \item Input: A secret key, $k$, the identifier, $\text{id}$, of the file, and a source block or encoded block, $\mathbf{e} \in \mathbb{F}^{n+m}_q$.
  \item Output: Tag $t$ for $\mathbf{e}$.
  \end{itemize}
\item $\cbn((\mathbf{e}_1,t_1,\alpha_1), \cdots, (\mathbf{e}_\ell,t_\ell,\alpha_\ell))$:
  \begin{itemize}
  \item Input: $\ell$ blocks, $\vct{e}_1, \cdots, \vct{e}_\ell$, their tags, $t_1, \cdots, t_\ell$, under key $k$, and their coefficients, $\alpha_1, \cdots, \alpha_\ell \in \mathbb{F}_q$.
  \item Output: Tag $t$ for block $\vct{e} \overset{\text{def}}{=} \sum_{i=1}^\ell \alpha_i\,\vct{e}_i$.
  \end{itemize}
\item $\vrf(k, \text{id}, \vct{e}, t)$:
  \begin{itemize}
  \item Input: A secret key, $k$,  the identifier, $\text{id}$, of the file,  a block, $\mathbf{e} \in \mathbb{F}^{n+m}_q$, and its tag, $t$.
  \item Output: 0 (reject) or 1 (accept).
  \end{itemize}
\end{itemize}
Also, the scheme must satisfy the following correctness requirement:\\
Let $t = \cbn((\mathbf{e}_1,t_1,\alpha_1), \cdots, (\mathbf{e}_\ell,t_\ell,\alpha_\ell)) $, then $\vrf \left( k, \text{id}, \sum_{i=1}^\ell \alpha_i \vct{e}_i, t \right) = 1$.

Note that the homomorphic property of the MAC scheme, or the existence of $\cbn$, which does not exist in regular MAC schemes, such as $\HMac$, ensures that multiple blocks can be audit at the bandwidth and verification computation cost of a single block. 

{\flushleft \bf Construction.} $\SMac$ consists of a triplet of algorithms: $\mac$, $\cbn$, and $\vrf$. The construction of $\SMac$ uses a pseudo-random function (PRF) $F_1: \mathcal{K}_1 \times (\mathcal{I} \times [1,n+m]) \rightarrow \mathbb{F}_q$, where $\mathcal{K}_1$ is the PRF key domain and $\mathcal{I}$ is the file identifier domain.

\begin{itemize}
\item $\mac(k, \iden, \vct{e}) \rightarrow t$: The MAC tag $t \in \eff{}{q}$ of a source block or encoded block, denoted by $\vct{e} \in \eff{n+m}{q}$, under key $k$, can be computed by the following steps:\\
-- $\vct{r}  \leftarrow (F_1 (k,  \iden, 1), \cdots, F_1 (k,  \iden, n+m))$ .\\
-- $t \leftarrow \vct{e} \cdot \vct{r} \in \mathbb{F}_q$ .

\item $\cbn((\mathbf{e}_1,t_1,\alpha_1), \cdots, (\mathbf{e}_\ell,t_\ell,\alpha_\ell)) \rightarrow t$: The tag $t \in \eff{}{q}$ of $\vct{e} \triangleq \sum_{i=1}^\ell \alpha_i \, \vct{e}_i \in \eff{n+m}{q}$ is computed as follows:\\
-- $t \leftarrow \sum_{i=1}^\ell \alpha_i \, t_i \in \mathbb{F}_q$ .

\item $\vrf(k, \iden, \vct{e}, t) \rightarrow \{0, 1\}$: To verify if $t$ is a valid tag of $\vct{e}$ under key $k$, we do the following:\\
-- $\vct{r}  \leftarrow (F_1 (k,  \iden, 1), \cdots, F_1 (k,  \iden, n+m))$ .\\
-- $t' \leftarrow \vct{e} \cdot \vct{r}$ .\\
-- If $t' = t$, output 1 (accept); otherwise, output 0 (reject).
\end{itemize}

\begin{lemma}[Theorem 1 in \cite{LeLocate2010}]\label{lemma:SpaceMac}
Assume that $F_1$ is a secure PRF. For any fixed $q$, $n$, $m$, {\SMac} is a secure $(q, n, m)$ homomorphic MAC scheme.
\end{lemma}

We refer the reader to \cite{LeLocate2010} for the security game and proof of \SMac. We provide security proof of $\SMac$ when used in {\ncaudit} in Section \ref{audit-subsec:possession}. If the user computes the verification tags for the source blocks using the {\mac} algorithm of {\SMac}, then the storage node can compute a valid MAC tag for any encoded block using the {\cbn} algorithm. The security of {\SMac} guarantees that if a block, $\vct{e}'$, is not a linear combination of the source blocks, then the storage node can only forge a valid MAC tag for $\vct{e}'$ with probability $\frac{1}{q}$. The security when using $\ell$ tags is improved to $\frac{1}{q^\ell}$. For clarity, we focus on a single file $\mathcal{F}$ and thus omit the file identifier {\iden} used by the above three algorithms in our subsequent discussion.

\subsection{The Random Linear Encryption: \ncrypt}
\label{audit-subsec:ncrypt}

To protect the privacy of the response file block, we need to encrypt it. The encryption, however, needs to still allow for the verification of the block. To this end, we design a novel encryption scheme that is compatible with {\SMac}, called {\ncrypt}. In particular, {\ncrypt} will protect $n-2$ elements of the response block while still allowing {\SMac} integrity checking. The remaining 2 elements are random padded elements. These 2 elements are needed to guarantee the security of the schemes, as we will show in the construction and proofs of {\ncrypt} and {\SMac}\footnote{In particular, the 2 random padded elements is to control the number of equations in the system of equations $\Pi_1$ and $\Pi_2$ described in the proofs of Theorems \ref{thm:NCrypt} and \ref{thm:PossessionSpaceMac}, respectively. Intuitively, these 2 random elements are needed to compensate for the extra information learned by the adversary in {\ncrypt} (the element $p$ as part of the ciphertext) and in {\SMac} (the equations related to $\bar{\vct{r}}$).}.

Let $\bar{\vct{x}}$ denote the vector formed by the first $n-2$ elements of a vector $\vct{x}$. The construction of $\ncrypt$ uses two PRFs: $F_2: \mathcal{K}_2 \times ([1,n-1] \times [1,n-2]) \rightarrow \mathbb{F}_q$ and $F_3: \mathcal{K}_2 \times (\{0,1\}^\lambda \times [1,n-1]) \rightarrow \mathbb{F}_q$, where $\mathcal{K}_2$ is a PRF key domain. {\ncrypt} consists of three probabilistic, polynomial time algorithms:

\begin{itemize}

\item $\setup(k, \bar{\vct{r}}) \rightarrow (p_1, \cdots, p_{n-1})$: This algorithm is run by the user to setup the encryption scheme. It takes as input a secret key $k$ and a vector $\bar{\vct{r}} \neq \vct{0}, \bar{\vct{r}} \in \eff{n-2}{q}$. It outputs $n-1$ elements in $\eff{}{q}\,$, which are called {\em auxiliary elements} and are used by the encryption. The details are as follows:\\ 
-- Compute $\bar{\vct{p}}_i \leftarrow ( F_2 (k, i, 1), \cdots, F_2 (k, i, n-2) ) \in \eff{n-2}{q}$, for $i \in [1,n-1]$.\\
-- Compute $p_i \leftarrow \bar{\vct{r}} \cdot \bar{\vct{p}}_i \in \eff{}{q}$, for $i \in [1,n-1]$.

\item $\enc(k, \bar{\vct{e}}, (p_1, \cdots, p_{n-1}) ) \rightarrow \langle \bar{\vct{c}}, (r, p) \rangle$: This algorithm is run by the storage node to encrypt the $n-2$ first elements of the aggregated response block. It takes as input a secret key, $k$, vector formed by the first $n-2$ elements of the response block, $\bar{\vct{e}}$, and the auxiliary elements, $p_1, \cdots, p_{n-1}$. It computes the encryption, $\langle \bar{\vct{c}}, (r, p) \rangle$, of $\bar{\vct{e}}$ as follows:\\
-- Compute $\bar{\vct{p}}_i, i \in [1,n-1]$, using key $k$ as in \setup.\\
-- Choose $r$ uniformly at random: $r \overset{R}{\leftarrow} \{0, 1\}^\lambda$.\\
-- Compute the {\em masking coefficients}: $\beta_i \leftarrow F_3 (k, r, i) \in \eff{}{q}, \text{for } i \in [1,n-1]\,.$\\
-- Compute the {\em masking vector}: $\bar{\vct{m}} \leftarrow \sum_{i=1}^{n-1} \beta_{i}\,\bar{\vct{p}}_i \in \eff{n-2}{q}\,.$\\
-- Compute $\bar{\vct{c}} \leftarrow \bar{\vct{e}} + \bar{\vct{m}} \in \eff{n-2}{q}$.\\
-- Compute $p \leftarrow \sum_{i=1}^{n-1} \beta_i \, p_i \in \eff{}{q}\,.$

In essence, the data is masked with a randomly chosen vector $\bar{\vct{m}} \in \lspan{ \bar{\vct{p}}_1, \cdots, \bar{\vct{p}}_{n-1} }$.

\item $\dec(k, \langle \bar{\vct{c}}, (r, p) \rangle) \rightarrow \bar{\vct{e}}$: This algorithm takes as input a secret key, $k$, and the cipher text, $\langle \bar{\vct{c}}, (r, p) \rangle$. The decryption is done as follows:\\
-- Compute $\bar{\vct{p}}_i, i \in [1,n-1]$, using key $k$ as in \setup.\\
-- Compute $\beta_i \leftarrow F_3 (k, r, i) \in \eff{}{q}$, for $i \in [1,n-1]$.\\ 
-- Compute $\bar{\vct{m}} \leftarrow \sum_{i=1}^{n-1} \beta_{i}\,\bar{\vct{p}}_i \in \eff{n-2}{q}$.\\
-- Compute $\bar{\vct{e}} \leftarrow \bar{\vct{c}} - \bar{\vct{m}} \in \eff{n-2}{q}$.
\end{itemize}

\begin{theorem}\label{thm:NCrypt}
Assume that $F_2$ and $F_3$ are secure PRFs, then {\ncrypt} is a fixed-length private-key encryption scheme for messages of length $(n-2) \times \log_2 q$ that has indistinguishable encryptions under a chosen-plaintext attack.
\end{theorem}

\begin{IEEEproof}
Intuitively, the security of {\ncrypt} holds because $\bar{\vct{m}}$ looks completely random to an adversary who observes a ciphertext $ \langle \bar{\vct{c}}, (r, p) \rangle $ since it is computationally difficult for the adversary to compute $\beta_i$'s without knowing the secret key $k$. 

The proof follows  a textbook technique used to prove the security of Construction 3.24 in \cite{Katz2007}. We follow the notation in \cite{Katz2007}. Denote the CPA security experiment of an encryption scheme $\Pi = (\setup, \enc, \dec)$ and an adversary $\mathcal{A}$ by $\mathsf{PrivK}^\mathsf{cpa}_{\mathcal{A}, \Pi}$. The game is as follows:
\begin{itemize}
\item A key $k$ is chosen uniformly at random from $\{0,1\}^\lambda$.
\item The adversary $\mathcal{A}$ is given $\bar{\vct{r}}, p_1, \cdots, p_{n-1}$, and oracle access to $\enc_k$. $\mathcal{A}$ outputs a pair of messages $\bar{\vct{e}}_0$ and $\bar{\vct{e}}_1$, both are in $\eff{n-2}{q}$.
\item A random bit $b \leftarrow \{0,1\}$ is chosen, and then a ciphertext $c \leftarrow \enc(k, \bar{\vct{e}}_b, (p_1, \cdots, p_{n-1}))$ is computed and given to $\mathcal{A}$. We call $c$ the challenge ciphertext.
\item The adversary $\mathcal{A}$ continues to have oracle access to $\enc_k$, and outputs a bit $b'$.
\item The output of the experiment is defined to be 1 if $b' = b$, and 0 otherwise. In case $\mathsf{PrivK}^\mathsf{cpa}_{\mathcal{A}, \Pi} = 1$, we say that $\mathcal{A}$ succeeded.
\end{itemize}

Let $\Pi_1$ be an encryption scheme that is exactly the same as $\Pi$ except that a truly random function $f_2$ is used in place of $F_2$.  Let Adv$[\mathcal{B}, F_2]$ be the probability of an adversary $\mathcal{B}$ with similar runtime to $\mathcal{A}$ winning the PRF security game (can tell a pseudo-random function $F_2$ from a truly random function $f_2$). By the security of PRF, we have that $\text{Adv}[\mathcal{B}, F_2]$ is negligible in $\lambda$ and it can be shown that (details are provided in the proof of Construction 3.24 in \cite{Katz2007})
\begin{equation}\label{audit-eq:PRF2}
\text{Adv}[\mathcal{B}, F_2] = | \text{Pr}[\mathsf{PrivK}^\mathsf{cpa}_{\mathcal{A}, \Pi} = 1] - \text{Pr}[\mathsf{PrivK}^\mathsf{cpa}_{\mathcal{A}, \Pi_1} = 1] |\,.
\end{equation}

Similarly, let $\Pi_2$ be an encryption scheme that is exactly the same as $\Pi_1$ except that a truly random function $f_3$ is used in place of $F_3$.  Let Adv$[\mathcal{C}, F_3]$ be the probability of an adversary $\mathcal{C}$ with similar runtime to $\mathcal{A}$ winning the PRF security game. Similar to the above, by the security of PRF, we have that $\text{Adv}[\mathcal{B}, F_3]$ is negligible in $\lambda$ and
\begin{equation}\label{audit-eq:PRF3}
\text{Adv}[\mathcal{C}, F_3] = | \text{Pr}[\mathsf{PrivK}^\mathsf{cpa}_{\mathcal{A}, \Pi_1} = 1] - \text{Pr}[\mathsf{PrivK}^\mathsf{cpa}_{\mathcal{A}, \Pi_2} = 1] |\,.
\end{equation}

We claim that for every adversary $\mathcal{A}$ that makes at most $g(\lambda)$ queries to its encryption oracle, where $g$ is a polynomial function, we have
\begin{align}{\label{audit-eq:CPA}}
\text{Pr}\left[ \mathsf{PrivK}^\mathsf{cpa}_{\mathcal{A}, \Pi_2} = 1 \right] \leq \frac{1}{2} + \frac{g(\lambda)}{2^\lambda}\,.
\end{align}

Let $r_c$ denote the random string used when generating the challenge ciphertext, which is of the form $\langle \bar{\vct{c}}, (r_c, p) \rangle$ (by encrypting $\bar{\vct{e}}_b$). There are two cases:

{\em (a) $r_c$ is never used by the oracle in the encryption algorithm to produce ciphertext to answer any of $\mathcal{A}$'s queries:} In the following, we will show that each element of {\em any} plaintext $\bar{\vct{e}}$ is masked with a uniformly random value, thus the adversary will not be able to tell which message ($\bar{\vct{e}}_0$ or $\bar{\vct{e}}_1$) was encrypted, as in the case of one-time pad.

Parse $\bar{\vct{e}}$ as $(e^{(1)}, \cdots, e^{(n-2)})$, $\bar{\vct{m}}$ as $(m^{(1)}, \cdots, m^{(n-2)})$, and $\bar{\vct{p}}_i$ as $(p^{(1)}_i, \cdots, p^{(n-2)}_i)$. From a ciphertext returned from an oracle query of $\bar{\vct{e}}$, the adversary can construct the following system of equations $\Pi_1$ by subtracting the query plaintext from the ciphertext:
\[
(\Pi_1) \quad
\begin{cases}
\beta_1 \, p^{(1)}_1 + \cdots + \beta_{n-1} \, p^{(1)}_{n-1} &= m^{(1)}\\
\cdots\\
\beta_1 \, p^{(n-2)}_1 + \cdots + \beta_{n-1} \, p^{(n-2)}_{n-1} &= m^{(n-2)}\\
\beta_1 p_1 + \cdots + \beta_{n-1} p_{n-1} &= p
\end{cases}
\,.
\]

Note that $p^{(j)}_i$ are not all zeros w.h.p. since they are chosen uniformly at random from $\eff{}{q}$ by $f_2$.  Let $\beta_i$ be unknowns, $i \in [1,n-1]$. The above system of $n-1$ linear equations is consistent regardless of the values of $m^{(j)}$'s since the rank of the coefficient matrix is at most $n-1$, which is the number of unknowns. Let $s$ be the rank of the coefficient matrix. Now for any $w \in [1,n-2]$, assume that all $m^{(j)},  j \neq w, j \in [1,n-2]$, are fixed. Then $m^{(w)}$ still can take any value in \eff{}{q} equally likely because (i) for any value of $m^{(w)}$, there is the same number of solutions, which is $q^{n - 1 - s}$, and (ii) $\beta_j$ are chosen uniformly at random from $\eff{}{q}$ (as a truly random function $f_3$ is used in place of $F_3$). Thus, each element of the plaintext, $e^{(w)}$, is masked with a uniformly random value, $m^{(w)}$, independent of other masking elements $m^{j \neq w}, j \in [1, n-2]$. Therefore, the probability that $\mathcal{A}$ outputs $b' = b$ is exactly 1/2, as in the case of the one-time pad. 

{\em (b)  $r_c$ is used by the oracle to answer at least one of $\mathcal{A}$'s queries:} In this case, $\mathcal{A}$ may easily determine which of its messages was encrypted. This is because whenever the oracle returns a ciphertext, $\langle \bar{\vct{c}}, (r, p) \rangle$, it learns the masking vector $\bar{\vct{m}}$ associated with $r$ as $\bar{\vct{m}} = \bar{\vct{c}} - \bar{\vct{e}}$. Thus, by leveraging the corresponding $\bar{\vct{m}}$ of $r_c$, the adversary can tell if $\bar{\vct{e}}_0$ or $\bar{\vct{e}}_1$ was encrypted by actually decrypting the challenge response. Since $\mathcal{A}$ makes at most $g(\lambda)$ queries, and $r$ is chosen uniformly at random, the probability of this event is at most $g(\lambda) / 2^\lambda$.

Equation (\ref{audit-eq:CPA}) follows from (a) and (b). Equations (\ref{audit-eq:PRF2}), (\ref{audit-eq:PRF3}), and (\ref{audit-eq:CPA}) show that \[\text{Pr}\left[ \mathsf{PrivK}^\mathsf{cpa}_{\mathcal{A}, \Pi} = 1 \right] \leq \frac{1}{2} + \frac{g(\lambda)}{2^\lambda} + \epsilon(\lambda)\,,\]
where $\epsilon$ is a cryptographically negligible function in $\lambda$. This completes the proof.
\end{IEEEproof}

\subsection{The Privacy-Preserving Auditing Scheme: \ncaudit}
\label{audit-subsec:audit-auditing}

Now we are ready to describe our symmetric-key based auditing protocol, called \ncaudit. In particular, \ncaudit is built from a novel combination of {\SMac} and {\ncrypt} as follows:

{\flushleft \em Setup phase:}
\begin{itemize}
\item The user divides the file into $m$ blocks of size $n-2$ instead of $n$ and pads to each block two random elements in $\eff{}{q}$. This is necessary as {\ncrypt} encrypts only the first $n-2$ elements. We still denote each padded block with its coding coefficients by $\vct{b}_i, i \in [1,m]$.

\item The user runs {\keygen} to generate MAC verification key, $k_v$, and encryption key, $k_e$:\\
-- $\keygen(1^\lambda) \rightarrow (k_e, k_v)$: $k_e \overset{R}{\leftarrow} \{0, 1\}^\lambda, k_v \overset{R}{\leftarrow} \{0, 1\}^\lambda$.

\item The user then setups the encryption scheme by computing the auxiliary elements, $p_1, \cdots, p_{n-1}$:\\
-- $\bar{\vct{r}} \leftarrow (F_1 (k_v, 1), \cdots, F_1 (k_v, n-2))$.\\
-- $(p_1, \cdots, p_{n-1}) \leftarrow \setup(k_e,\bar{\vct{r}})$.

\item Afterward, the user computes a tag for each source block $\vct{b}_i$ using the {\mac} algorithm of {\SMac}:\\ 
-- $t_{\vct{b}_i} = \mac(k_v, \vct{b}_i)$.

\item The user computes MAC tags of encoded blocks using the {\cbn} algorithm of {\SMac}. Assume $\vct{e} = \sum_{i=1}^{m} \alpha_i \, \vct{b}_i$, then its tag is computed as follows:\\
-- $\taggen(\vct{e}, k_v) \rightarrow t_{\vct{e}} = \sum_{i=1}^{m} \alpha_i \, t_{\vct{b}_i}$.

\item Finally, the user sends the encoded blocks, $\vct{e}_1, \cdots, \vct{e}_M$, their tags, $t_{\vct{e}_1}, \cdots, t_{\vct{e}_M}$, the auxiliary elements, $p_1, \cdots, p_{n-1}$, and the encryption key, $k_e$, to the storage node. The user also sends the coding coefficients, $\aug{\vct{e}_1}, \cdots, \aug{\vct{e}_M}$, and the MAC key, $k_v$, to the TPA. We assume that the user uses private and authentic channels to send $k_v$ and $k_e$\footnote{Exchanging secret keys, in particular, and establishing secure and authentic channels, in general, could be done with the support of a public key infrastructure (PKI). This is an important, well studied problem in the cryptography community and is orthogonal to this work.}.
The user then keeps the coding coefficients and the keys but delete all other data.  

\end{itemize}

Note that maintaining coding coefficients is necessary for the repair process and is an inherent characteristic of NC storage systems. The overhead of storing the coefficients is negligible compared to the outsource data and could be  constant for practical purposes (see Section \ref{audit-subsec:storageOverhead}). If the user outsources the management of the nodes to a third party, such as a proxy as in NCCloud \cite{Hu2012}, then he/she does not need to store the coding coefficients. However, in this case, the proxy must be trusted.

{\flushleft \em Audit phase:}
\begin{itemize}
\item The TPA chooses a set of indexes of blocks to be audited, $\mathcal{I} \subseteq [1,M]$, and chooses the coefficients for these blocks uniformly at random: $\alpha_i \overset{R}{\leftarrow} \eff{}{q}, i \in \mathcal{I}$. The challenge includes the indexes of the blocks and their corresponding coefficients:\\
-- Prepare $\chal = \{ (i, \alpha_i) \,|\, i \in \mathcal{I} \}$.

\item {\genproof} run by the storage node to generate the proof of storage, $V$, is implemented as follows:\\
-- Compute the aggregated block: $\hat{\vct{e}} = \sum_{i \in \mathcal{I}} \alpha_i \, \hat{\vct{e}}_i$. Parse $\hat{\vct{e}}$ as $(\bar{\vct{e}}, e^{(n-1)}, e^{(n)})$.\\
-- Compute the aggregated tag: $t = \sum_{i= \in \mathcal{I}} \alpha_i \, t_{\vct{e}_i}$.\\
-- Encrypt the response block: $\langle \bar{\vct{c}}, (r, p) \rangle \leftarrow \enc(k_e, \bar{\vct{e}}, (p_1, \cdots, p_{n-1}))$.\\
The node then sends $V = (\langle \bar{\vct{c}}, (r, p) \rangle, e^{(n-1)}, e^{(n)}, t)$ back to the TPA.

\item $\verifyproof$ run by the TPA to verify the proof $V$ is implemented as follows:\\
-- Compute coefficients of $\hat{\vct{e}}$: $\aug{\vct{e}} = \sum_{i \in \mathcal{I}} \alpha_i \, \aug{\vct{e}_i}$.\\
-- Let $\vct{c} = (\bar{\vct{c}} \, | \, e^{(n-1)} \,|\, e^{(n)} \, | \, \aug{\vct{e}})$, where ``$|$'' denotes augmentation.\\
\hspace*{2mm} Return result of $\vrf(k_v, \vct{c}, t + p)$.
\end{itemize}

{\flushleft \bf Correctness.} The correctness of {\ncaudit}, \ie, if the file is correct then the algorithm will accept the proof, is guaranteed by the following Lemma \ref{thm:correctness}. And its security, \ie, if there is corruption then the algorithm will reject the proof, is proved in Section \ref{sec:audit-security}.
\begin{lemma}\label{thm:correctness}
If the storage node follows {\ncaudit} and computes the aggregated response block using uncorrupted blocks, then the TPA will accept the proof.
\end{lemma}

\begin{IEEEproof}
Let  $\vct{r} = (F_1 (k_v,  1), \cdots, F_1 (k_v, n+m))$.
Note that 
\begin{align*}
\vct{c} &= (\bar{\vct{c}}\,|\, e^{(n-1)} \,|\, e^{(n)}\,|\, \aug{\vct{e}})\\
~ &= ((\bar{\vct{e}} + \bar{\vct{m}})\,|\, e^{(n-1)} \,|\, e^{(n)} \,|\, \aug{\vct{e}}) = \vct{e} + (\bar{\vct{m}}\,|\, 0, \cdots, 0)\,. 
\end{align*}

Thus, in the {\vrf}, 
\begin{align*}
t' &= \vct{c} \cdot \vct{r} = \vct{e} \cdot \vct{r} + \bar{\vct{m}} \cdot \bar{\vct{r}}\\
~ &= t + \sum_{i=1}^{n-1} \beta_i \, \bar{\vct{p}}_i \cdot \bar{\vct{r}} = t + \sum_{i=1}^{n-1} \beta_i \, p_i = t + p\,.
\end{align*}
Therefore, {\vrf} returns 1. Hence, the TPA accepts the proof.
\end{IEEEproof}

\section{Support for Node Repair}
\label{sec:audit-repair}


\label{audit-subsec:repair}
When there is a node failure, the user creates a new node to replace this node. Based on the coding coefficients of the coded blocks at the remaining healthy nodes, the user instructs the healthy nodes to send appropriate coded blocks to the new node. The new node then linearly combines them, according to the user instruction, to construct its own coded blocks. This new node may construct the same coded blocks that the failed node had ({\em exact repair}), or completely different coded blocks that still preserve the same level of reliability ({\em functional repair}) \cite{Dimakis2011}. In the example given in Fig. \ref{fig:repair}, the user instructs the first three storage nodes to send coded blocks to exactly repair the fourth node. 

Formally, for each healthy node, $N_i, i = 1, \cdots, P$, recall that it needs to send $Q$ encoded repair blocks to the new node. Let $(\vct{e}_{i,1}, \cdots, \vct{e}_{i,M})$ be the encoded blocks currently stored on $N_i$. For $j = 1, \cdots, Q$, the user sends a set of {\em repair coding coefficients} $(\gamma_{i,j,1}, \cdots, \gamma_{i,j,M})$ to $N_i$. This node then uses these coefficients to compute the repair blocks, $\vct{g}_{i,j} = \sum_{k=1}^M \gamma_{i,j,k} \, \vct{e}_{i,k}$, to send to the new node. The new node will receive $P \times Q$ repair blocks, $\vct{g}_{i,j}$, from the healthy nodes. It uses them to reconstruct the encoded blocks, $\vct{h}_1, \cdots, \vct{h}_M$, that it needs to store. For $k = 1, \cdots, M$, the user sends a set of $P \times Q$ {\em reconstruction coding coefficients}, $(\theta_{i,j,k}, \cdots \theta_{P,Q,k})$, to the new node to instruct its reconstruction. The new node then reconstructs $\vct{h}_k = \sum_{i=1}^{P} \sum_{j=1}^{Q} \theta_{i,j,k} \, \vct{g}_{i,j}$. Note that the coding coefficients $\gamma$'s and $\theta$'s are dependent on the repairing scheme.

Using {\ncaudit}, the verification tags of the newly constructed blocks, $\vct{h}_k$, at the new node do not need to be computed by the user. In particular, the healthy nodes can send along the verification tags of the repair blocks, $\vct{g}_{i,j}$, that they send to the new node, where the tags of $\vct{g}_{i,j}$ can be computed using the $\cbn$ algorithm of $\SMac$ on the tags of $\vct{e}_{i,k}$. The new node then can also use {\cbn} on the tags of $\vct{g}_{i,j}$ to generate tags of $\vct{h}_k$. Finally, the user sends the coding coefficients of the coded blocks at the newly constructed node, $\aug{\vct{h}_k}$ (dependent on the repair scheme), to the TPA so that it can audit this new node.

Consequently, with {\ncaudit}, there is negligible cost to the user when repairing a failed node, in terms of both bandwidth and computation of verification metadata. In particular, the user does not need to download data, \ie, $\vct{e}_{i,k}$, and the user also does not need to compute the tags, \ie, runs {\mac} on $\vct{h}_k$. This stands in stark contrast with the prior integrity checking scheme for NC-based storage \cite{Chen2010}, which requires the user to download many data blocks (equal to the repair bandwidth) and compute security metadata for the newly coded blocks him/herself. 

Last but not least, since the TPA audits the new node based on the new set of coefficients, a malicious node cannot carry out a {\em replay attack} \cite{Chen2010} (discussed in Section \ref{audit-subsec:nc_based_check}); otherwise, it will not pass the audit because the {\SMac} tags are computed on both the data and coefficients. Here we assume that the healthy remaining nodes send valid data and tags to the new node. If there is a malicious node that sends corrupted data or tags, the storage systems is considered polluted. Dealing with pollution attacks is out of the scope of this paper; we refer the reader to previous work, including our own, which explicitly combats pollution attacks \cite{Gkantsidis2006, Agrawal2009, Boneh2009, Agrawal2010, Li2010, LeJSAC2011, LeTESLA2011, LeInter2012, Buttyan2010}.

\section{Security Analysis}
\label{sec:audit-security}

\subsection{Data Possession Guarantee}
\label{audit-subsec:possession}

When using $\SMac$ in $\ncaudit$, some information about the vector $\vct{r}$ in the $\SMac$ construction is available to the adversary. In particular, the storage node knows the following $n-1$ equations: $\bar{\vct{p}}_i \cdot \bar{\vct{r}} = p_i\,, i \in [1,n-1]$. The following theorem states that even when these $n-1$ equations are exposed, $\SMac$ is still a secure homomorphic MAC, \ie, any corruption will be detected w.h.p.

\begin{theorem}\label{thm:PossessionSpaceMac}
Assume that $F_1$ is a secure PRF. For any fixed $q$, $n$, $m$, assume that a probabilistic polynomial time adversary $\mathcal{A}$ knows any $n-1$ linearly independent vectors, $\bar{\vct{p}}_1, \cdots, \bar{\vct{p}}_{n-1}$, and any $n-1$ constants, $p_1, \cdots, p_{n-1}$, such that  $\bar{\vct{p}}_i \cdot \bar{\vct{r}} = p_i$, where $\vct{r}$ is used in the construction of $\SMac$. The probability that $\mathcal{A}$ wins the $\SMac$ security game, denoted by $\text{\em Adv}[\mathcal{A}, \SMac]$, is at most
\[ \text{\em PRF-Adv}[\mathcal{B}, F_1] + \frac{1}{q}\,,\]
where {\em PRF-Adv}$[\mathcal{B}, F_1]$ is the probability of an adversary $\mathcal{B}$ with similar runtime to $\mathcal{A}$ winning the PRF security game.
\end{theorem}

\begin{IEEEproof}
The security game, called the Attack Game 1, of $\SMac$ involves a challenger $\mathcal{C}$ and an adversary $\mathcal{A}$, and is as follows:
\begin{itemize}
\item {\em Setup.} $\mathcal{C}$ generates a random key $k \overset{\text{R}}{\leftarrow} \mathcal{K}$
\item {\em Queries.} $\mathcal{A}$ adaptively queries $\mathcal{C}$, where each query is of the form $(\text{id}, \vct{y})$. For each query,  $\mathcal{C}$ replies to $\mathcal{A}$ with the corresponding tag $t \leftarrow \mathsf{Mac}(k, \text{id}, \vct{y})$.
\item {\em Output.}  $\mathcal{A}$ eventually outputs a tuple $(\text{id}^*, \vct{y}^*, t^*)$.
\end{itemize}
Up to the time $\mathcal{A}$ outputs, it has queried $\mathcal{C}$ multiple times. Let $l$ denote the number of times $\mathcal{A}$ queried $\mathcal{C}$ using $\text{id}^*$ and get tags for $l$ vectors, $\vct{y}^*_1, \cdots, \vct{y}^*_l$, of these queries. We consider that the adversary wins the security game if and only if
\begin{itemize}
\item $(y_*^{(n+1)}, \cdots, y_*^{(n+m)}) \neq \vct{0}$ (trivial forge otherwise),
\item $\mathsf{Verify} (k, \text{id}^*, \vct{y}^*, t^*) = 1$, and
\item $\vct{y}^* \notin \mathsf{span}(\vct{y}^*_1, \cdots, \vct{y}^*_l).$
\end{itemize}

Here, we prove Theorem \ref{thm:PossessionSpaceMac} with respect to a slightly different security game, called Attack Game 2. This Attack Game 2 is similar to Attack Game 1, except that in the {\em Queries} phase, for each distinct id, the space spanned by the vectors used in the queries has dimension at most $m$. This Attack Game 2 is stricter but better fits the reality: since the dimension of the source space $\Pi$ is only $m$, the adversary must only learn tags of vectors in spaces having dimensions at most $m$.

Now the proof is done by using a sequence of games denoted Game 0 and Game 1. Let $W_0$ and $W_1$ denote the events that $\mathcal{A}$ wins the homomorphic MAC security in Game 0 and Game 1, respectively. Game 0 is identical to Attack Game 2 applied to the scheme $\SMac$. Hence,
\begin{equation}\label{audit-eq:1}
\text{Pr}[W_0] = \text{Adv}[\mathcal{A}, \SMac]
\end{equation}
Game 1 is identical to Game 0 except that the challenger $\mathcal{C}$ computes $\vct{r} \leftarrow (r_1, \cdots, r_{n+m})$, where $r_i$ is chosen uniformly at random from $\eff{}{q}$: $r_i \overset{\text{R}}{\leftarrow} \mathbb{F}_q$ instead of $r_i \leftarrow F(k, \text{id}, i)$, and everything else remains the same. Then, there exists a PRF adversary $\mathcal{B}$ such that
\begin{equation}\label{audit-eq:2}
|\text{Pr}[W_0] - \text{Pr}[W_1]| = \text{PRF-Adv}[\mathcal{B}, F]
\end{equation}

The complete challenger in Game 1 works as follows:

{\flushleft \emph{Queries.}} $\mathcal{A}$ adaptively queries $\mathcal{C}$, where each query is of the form $(\text{id}, \vct{y})$. If id is already used in $m$ previous query, $\mathcal{C}$ discards the query. Otherwise, $\mathcal{C}$ replies to query $i$ of $\mathcal{A}$ as follows:\\
\hspace*{0.3 cm} if id is never used in any of the previous queries:\\
\hspace*{0.6 cm} $\vct{r}_i := (r^i_1, \cdots, r^i_{n+m})$, where $r^i_j \overset{\text{R}}{\leftarrow} \mathbb{F}_q, j \in [1,n+m]$\\
\hspace*{0.3 cm} else:\\
\hspace*{0.6 cm} $\vct{r}_i$ := the one used in the previous response\\
\hspace*{0.3 cm} send $t := \vct{y}_i \cdot \vct{r}_i$ to $\mathcal{A}$

{\flushleft \emph{Output.}}  $\mathcal{A}$ eventually outputs a tuple $(\text{id}^*, \vct{y}^*, t^*)$. When $\vct{y}^*$ does not equal $\vct{0}$, to determine if $\mathcal{A}$ wins the game, we compute\\
\hspace*{0.3 cm} if $\text{id}^* = \text{id}_i$ (for some $i$) then \hspace*{0.5 cm} // case (i)\\
\hspace*{0.6 cm} set $\vct{r}^* := \vct{r}_i$\\
\hspace*{0.3 cm} else \hspace*{5 cm} // case (ii)\\
\hspace*{0.6 cm} set $\vct{r}^* := (r^*_1, \cdots, r^*_{n+m})$, where $r^*_i \overset{\text{R}}{\leftarrow} \mathbb{F}_q, i \in [1,n+m]$\\
Let $l$ denote the number of times $\mathcal{A}$ queried $\mathcal{C}$ using $\text{id}^*$ and get tags for $l$ vectors, $\vct{y}^*_1, \cdots, \vct{y}^*_l$, of these queries. The adversary wins the game, \emph{i.e.}, event $W_1$ happens, if and only if
\begin{align}
~&t^* = \vct{y}^* \cdot \vct{r}^*\,,\text{ and}\label{audit-eq:3}\\
~&\vct{y}^* \notin \mathsf{span}(\vct{y}^*_1, \cdots, \vct{y}^*_l)\label{audit-eq:4}\,.
\end{align}

Subsequently, we will show that Pr[$W_1$] = $\frac{1}{q}$. Let $T$ be the event that $\mathcal{A}$ outputs a tuple with a completely new $\text{id}^*$, \emph{i.e.}, $\mathcal{A}$ never made queries using $\text{id}^*$ before.

$\bullet$ When T happens, \emph{i.e.}, in case (ii), since $r^*_i\,$'s are indistinguishable from random values and $(y_*^{(n+1)}, \cdots, y_*^{(n+m)}) \neq \vct{0}$, the right hand side of equation (\ref{audit-eq:3}) is a completely random value in $\mathbb{F}_q$. Thus,
\begin{equation}\label{audit-eq:5}
\text{Pr}[W_1 \wedge T] = \frac{1}{q}\,\text{Pr}[T]\,.
\end{equation}

$\bullet$ When T does not happen, \emph{i.e.}, in case (i): $\vct{r}^*$ of equation (\ref{audit-eq:3}) equals $\vct{r}_i$ for some $i$, and $\vct{r}^*$ has been used to generate tags for vectors $\vct{y}^*_1, \cdots, \vct{y}^*_l$.  In this case, we proceed by showing that for a fixed $\vct{y}^*$, $t^*$ looks indistinguishable from a random value in $\mathbb{F}_q$. The given prior knowledge, the queries, and the output form the following system of linear equations $\Pi_2$: 
\[
(\Pi_2) \quad
\begin{cases}
\bar{\vct{p}}_1 \cdot \bar{\vct{r}}^* = p_1\\
\cdots~\\
\bar{\vct{p}}_{n-1} \cdot \bar{\vct{r}}^* = p_{n-1}\\
\vct{y}^*_1 \cdot \vct{r}^* = t_{\vct{y}^*_1}\\
\cdots~\\
\vct{y}^*_l \cdot \vct{r}^* = t_{\vct{y}^*_l}\\
\vct{y}^* \cdot \vct{r}^* = t^*
\end{cases}
\,.
\]

Let the elements $r^*_i, i \in [1,n+m]$, of $\vct{r}^*$ be the unknowns of the system. The above system is consistent regardless of the value of $t^*$ because the coefficient matrix has rank at most $n+m$, which equals the number of unknowns. Let $d$ be the rank of the coefficient matrix, $d\leq n+m$. For a fixed $\vct{y}^*$, its valid tag $t^*$ could be any value in $\mathbb{F}_q$ equally likely because (i) for any value $t^*$, the solution space always has the same size $q^{n+m-d}$, and (ii) $r^*_i$'s are chosen uniformly at random from $\mathbb{F}_q$. As a result, the probability that the adversary chooses a correct $t^*$  is $1/q$. Thus,
\begin{equation}\label{audit-eq:6}
\text{Pr}[W_1 \wedge \neg T] = \frac{1}{q}\,\text{Pr}[\neg T]\,.
\end{equation}

$\bullet$ From equations (\ref{audit-eq:5}) and (\ref{audit-eq:6}), we have
\begin{equation}\label{audit-eq:7}
\text{Pr}[W_1] = \text{Pr}[W_1 \wedge T] + \text{Pr}[W_1 \wedge \neg T] = \frac{1}{q}\,.
\end{equation}

Equations  (\ref{audit-eq:1}), (\ref{audit-eq:2}), and  (\ref{audit-eq:7}) together prove the theorem.
\end{IEEEproof}


Now, we are ready to prove the data possession guarantee of {\ncaudit}.

\begin{lemma}\label{thm:possession}
With probability at least $1 - \frac{2}{q}$, the storage node can pass a check if and only if it possesses the blocks specified in the challenge of the check.
\end{lemma}

\begin{IEEEproof}
Lemma \ref{thm:correctness} shows that if the storage node possesses the data then it can pass the check. It remains to show that if the node passes the check then it possesses the corresponding blocks w.h.p. Let us prove the converse, \ie, if there are corrupted or missing blocks, the node will fail the check w.h.p. For simplicity, we assume that when responding to a challenge involving a block that no longer exists in the storage, the node replaces it with a block chosen uniformly at random in $\eff{n+m}{q}$. 

{\em Case (a) - The storage node is able to compute a correct response block even when some blocks are missing or corrupted:} Denote the correct, unencrypted aggregated block by $\vct{e}$, \ie, $\vct{e} = \sum_{i \in \mathcal{I}} \alpha_i \, \vct{e}_i $. Denote the data of the response block actually computed by the storage node by $\hat{\vct{a}}$ and denote $(\hat{\vct{a}} \, | \, \aug{\vct{e}} )$ by $\vct{a}$. If there is at least one error in the data of one of the block or there is at least one missing block, then $\Prob{\hat{\vct{a}} = \hat{\vct{e}}} \leq \frac{1}{q}$ because $\alpha$'s are chosen uniformly at random from $\eff{}{q}$. Note that $\vct{e}$ is in the source space: $\vct{e} \in \Pi$, thus if $\hat{\vct{a}} \neq \hat{\vct{e}}$ then $\vct{a} \notin \Pi$. Therefore, $\Prob{\vct{a} \in \Pi} = \Prob{\vct{a} = \vct{e}} \leq \frac{1}{q}$. 

{\em Case (b) - The storage node responds with an incorrect block:} The security of $\SMac$ from Theorem \ref{thm:PossessionSpaceMac} guarantees that the node can provide a valid tag of $\vct{a} \notin \Pi$ with probability at most $\frac{1}{q}$. Without loss of generality, we can ignore the encryption because if the node already knows a valid tag of $\vct{a}$, it can provide the correct encryption to pass the check. Meanwhile, if the node does not know a valid tag of $\vct{a}$, its chance of forging a valid tag for the cipher text $\vct{c}$ is still bounded by the security guarantee of $\SMac$, which is at most $\frac{1}{q}$. 

As a result, from cases (a) and (b), the probability of passing the check when there is error or missing block is at most $\frac{2}{q}$.
\end{IEEEproof}


Not only does {\ncaudit} provide detection in the presence of corrupted or missing blocks, it also ensures that the user can extract the data stored on the storage node just by collecting responses of the node from the checking protocol. This is also known as the {\em retrievability} property. We provide the proof of retrievability based on the theoretical framework of \cite{Bowers2009}, which is derived from \cite{Shacham2008} and \cite{Juels2007}. 

\begin{lemma}
Assume that the storage node responds correctly to a fraction, $1-\epsilon$, of the challenges uniformly, where $\epsilon < \frac{1}{2}$. The user can extract the encoded blocks stored on the node, $\vct{e}_1, \cdots, \vct{e}_M$, by performing $\gamma$ challenge-response interactions with the storage node with high probability (depending on $\gamma$, $\epsilon$, and $q$).
\end{lemma}

\begin{IEEEproof}
Lemma \ref{thm:possession} implies that if a node responds correctly to a fraction of challenge, then with probability at least $1-\frac{2}{q}$, the response block is a correct linear combination of the blocks stored at the node. For a challenge coefficient vector $(\alpha_1, \cdots, \alpha_M)$, the user can challenge the node using a number of constant-multiples of the vector, \eg, $(c\,\alpha_1, \cdots, c\,\alpha_M)$ for some constant $c$, to learn the responses (including incorrect responses), and then use majority decoding to learn the correct equation $\sum_{i=1}^M \alpha_i \vct{e}_i = \vct{d}$, where $\vct{d}$ is some constant vector. By collecting $M$ linearly independent equations of this form, the user can solve for $\vct{e}_1, \cdots, \vct{e}_M$ using Gaussian elimination.

Note that for a fixed $\epsilon < \frac{1}{2}$, the probability of learning one correct equation depends on both $q$ and the number of queries made using the multiples of the corresponding coefficient vector. For a fixed $q$, this probability can be made arbitrarily high by increasing the number of queries.
\end{IEEEproof}

\subsection{Privacy-Preserving Guarantee}
\label{audit-subsec:privacy}

{\ncrypt} provides the privacy guarantee of {\ncaudit}, which we stated in the following lemma.

\begin{lemma}
From the responses of the storage node, the TPA does not learn any information about the outsourced data, except for the information that could be derived from the MAC tag.
\end{lemma}

\begin{IEEEproof}
The claim is a direct consequence of Theorem \ref{thm:NCrypt} and the fact that the padding elements are chosen randomly.
\end{IEEEproof}


\section{Performance Evaluation}
\label{sec:audit-evaluation}

\subsection{Client Storage Overhead}
\label{audit-subsec:storageOverhead}

{\ncaudit} requires the user and the TPA to store the coding coefficients, which is in $O (m M N)$ space. The user needs the coefficients to carry out repairs while the TPA needs the coefficients to carry out audits. In any case, the overhead of $O (m M N)$ is orders of magnitude less than the outsourced data, which is in $O ( (n+m) M N )$ space; this is because $n \gg m$ for NC-based storage systems. In fact, in a practical NC storage cloud, the space necessary for storing the coding coefficients could be kept less than 160 B (\ie, constant storage) while being able to support arbitrary file size (by increasing the block size $n$, see Section 5.1 of NCCloud \cite{Hu2012}). Table \ref{tab:comparison} compares client storage overhead of {\ncaudit} and other state-of-the-art schemes \cite{QWang2009, CWang2010, Chen2010}.

\subsection{Bandwidth Overhead}
\label{audit-subsec:bandwidthOverhead}

{\flushleft \bf Integrity Checking.}\quad For each audit round, the major communication cost is the cost of sending the proof of possession from the storage node to the TPA, which is dominated by the size of the (encrypted) data bock. Thanks to homormophic property of {\SMac}, blocks in the challenge can be aggregated. We achieve similar bandwidth overhead compared to prior schemes for integrity checking of cloud data \cite{Shacham2008, QWang2009, Chen2010, CWang2010}. In particular, the proof of possession for multiple blocks contains only a single block (of size varying from 4 KB \cite{Ateniese2007} to 1.6 MB \cite{Chen2010}).


{\flushleft \bf Repairing.}\quad As discussed in Section \ref{sec:audit-repair}, when using {\ncaudit}, the user does not need to download any data block to repair a failed node. This stands in stark contrast with the state-of-the-art scheme for NC storage systems \cite{Chen2010}, where the user needs to download an amount of data equal to the repair bandwidth to setup integrity metadata for the new coded blocks him/herself.

{\flushleft \bf Encryption.}\quad The amount of additional bandwidth to support encryption is small. In particular, {\ncrypt} requires the storage node  to send with the encrypted block, $\bar{\vct{c}}$; the random value, $r$, of size $\lambda$ (typically 80 bits \cite{Ateniese2007}); the auxiliary tag, $p$, and the random padding elements, $e^{(n-1)}, e^{(n)}$, which are of size $\log_2 q$. These are negligible compared to the block size: $n \log_2 q$, \eg, 0.3\% for $q = 2^8, n=4 \times 2^{10}$ (4 KB block).

The bandwidth overhead of {\ncaudit} when compared to other schemes  \cite{QWang2009, CWang2010, Chen2010} are summarized in Table \ref{tab:comparison}.

\subsection{Computational Overhead}
\label{audit-subsec:computationOverhead}

\begin{table*}[t]

\begin{center}

{\footnotesize
\begin{tabular}{|l|l|l|l|l|l|}
\hline
~ & ~ & {\bf Wang 2009 \cite{QWang2009}} & {\bf Wang 2010 \cite{CWang2010}} & {\bf Chen 2010 \cite{Chen2010}} & {\bf \ncaudit}\\
\hline
\multirow{3}{*}{ \bf Features } & ~ & Public-Key Audit & Public-Key Audit & Private-Key Audit & Private-Key Audit\\
\cline{3-6}
  & ~ & {\em No} NC Repair & {\em No} NC Repair & NC Repair & {\em Efficient} NC Repair\\
\cline{3-6}
  & ~ & {\em No} Audit Privacy  & Audit Privacy  & {\em No} Audit Privacy  & Audit Privacy \\
\hline
\multirow{2}{*}{ \bf Client Storage } & Audit Overhead & $O(1)$ & $O(1)$ & $O(1)$ & $O(mMN)$\\
\cline{2-6}
  & Repair Overhead & N/A & N/A & $O(mMN)$ & $O(mMN)$\\
\hline
\multirow{3}{*}{ \bf Bandwidth } & Audit Overhead & 1 block & 1 block & 1 block & 1 block\\
\cline{2-6}
  & Repair  Overhead & N/A & N/A & repair bandwidth & 0*\\
\cline{2-6}
  & Enc.  Overhead & N/A & 0* & N/A & 0*\\
\hline
\multirow{4}{*}{ \bf Computation } & Security & \multicolumn{4}{|c|}{80-bit}\\
\cline{2-6}
  & Parameters & \multicolumn{4}{|c|}{300 blocks per challenge, 4 KB block size}\\
\cline{2-6}
  & Testbed Config. & \multicolumn{2}{|c|}{1.86 Ghz CPU, 2GB RAM} & \multicolumn{2}{|c|}{2.8 Ghz CPU, 32 GB RAM}\\
\cline{2-6}
  & Server Overhead & 270 ms & 273 ms & 3.19 ms & 4.69 ms\\
\cline{2-6}
  & Auditor Overhead & 491 ms & 493 ms & 2.76 s & 0.73 ms\\
\hline
\end{tabular}
}
\end{center}

\caption{Comparison of different remote data integrity checking schemes. 0* indicates no data block needs to be downloaded by the user to support the feature. N/A means not applicable due to the lack of support.}
\label{tab:comparison}
\vspace*{-20pt}
\end{table*}

We first analyze the cost of each operation in {\ncaudit} by the number of finite field multiplications involved, which is the dominating cost factor. We then present the cost of each operation from our real implementation in Java. We omit the cost of computing PRF values that do not take as input random seeds since they can be precomputed.

{\flushleft \bf Integrity Checking with Encryption:}

{\flushleft \em 1. Storage Node Overhead:} In {\ncaudit}, the cost to compute a proof of possession includes the cost to compute (i) the aggregated response block, $\bar{\vct{e}}$, (ii) the response tag, $t$, (iii) the masking vector, $\bar{\vct{m}}$, and the auxiliary element, $p$. The total cost is dominated by the cost to compute $\bar{\vct{e}}$ and $\bar{\vct{m}}$. $\bar{\vct{m}}$ can be precomputed in advance as it is independent of the challenge. Let $C$ be the average number of blocks specified in a challenge. The average cost to compute a response per challenge is $C \times n$ multiplications with a precomputation of $\bar{\vct{m}}$ and $C\times n + (n-2) \times (n-1)$ without.

{\flushleft \em 2. TPA Overhead:} In {\ncaudit}, verifying a proof of possession can be done very efficiently. In particular, the cost to verify include the time to (i) compute the coefficients of the response block and (ii) run the $\vrf$ of $\SMac$. Let $\ell$ be the number of tags used (to increase the security to $1/q^\ell$). The total cost is $C \times m + \ell \times (n+m)$ multiplications.

{\flushleft \bf Repairing:} 

As described in Section \ref{sec:audit-repair}, repairing a failed node does not incur any computation cost at the user side to maintain the security metadata of the auditing.

{\flushleft \bf Implementation:} 

We implement {\ncaudit} in Java to compare its performance with the state-of-the-art schemes \cite{QWang2009, CWang2010, Chen2010}. For a fair of comparison with \cite{QWang2009, CWang2010}, we use $q = 2^8$ and $\ell = 10$ to provide 80-bit security, and we also set block size to 4 KB ($n = 4 \times 2^{10}$), $m=500$, and the number of blocks indicated by a challenge to $C = 300$. We stress that the choice of parameters may be different in a practical NC storage system, \eg, in \cite{Hu2012}, a block size could be as big as 4 MB while the storage space taken by the coefficients could be kept below 160 B. We implement finite field multiplications in $\eff{}{2^8}$ by table look-ups and additions using XORs. 
We also precomputed values that do not depend on the challenges.

Table \ref{tab:comparison} compares the computational overhead of different remote data integrity checking schemes. The reported numbers for \cite{QWang2009} and \cite{CWang2010} are taken from \cite{CWang2010}. (The overhead of the scheme in \cite{QWang2009} is similar to the public-key based scheme in \cite{Shacham2008}.) We refer the reader to \cite{CWang2010} for the detailed setup. We implement the checking scheme in \cite{Chen2010} ourselves. For this scheme, we use AES with CBC mode from the Java {\em crypto} library to decrypt coding coefficients. We refer the reader to Appendix A in \cite{Chen2010} for the detailed description of this scheme. The number reported for {\ncaudit} and the scheme in \cite{Chen2010} are the average of 100 runs on a computer with 2.8 Ghz CPU and 32 GB RAM. We note that among the three schemes under comparison \cite{QWang2009, CWang2010, Chen2010}, the scheme in \cite{Chen2010} is the only one specifically designed for NC storage systems and thus supports NC repair.

Table \ref{tab:comparison} shows that {\ncaudit} manages to achieve very modest computational overhead. The computational overhead of {\ncaudit} is orders of magnitude smaller than those of \cite{QWang2009} and \cite{CWang2010}. This is due to the fact that {\ncaudit} is symmetric-key based while the schemes in \cite{CWang2010} and \cite{QWang2009} are public-key based and make heavily use of expensive bilinear mapping operations\footnote{Due to the fundamental difference: the use of expensive bilinear mapping operations in \cite{CWang2010, QWang2009}, we expect a similar gap (in order of magnitude) between the computational overhead of \cite{CWang2010, QWang2009} and that of {\ncaudit} when we run them on the same hardware.}. The scheme in \cite{Chen2010} achieves similar storage node's computational overhead to {\ncaudit} as it is also symmetric-key based. However, due to the cost of executing $C \times m = 150,000$ numbers of decryption for the coefficients, the computational overhead of the TPA of \cite{Chen2010} is much larger than that of {\ncaudit}, in the order of seconds as opposed to milliseconds.


\section{Conclusion}
\label{sec:audit-conclusion}

In this work, we propose {\ncaudit}, a cryptography-based remote data integrity checking scheme for NC-based storage systems. {\ncaudit} is based on a novel combination of an existing MAC scheme custom made for network coding, $\SMac$, and a novel CPA-secure encryption scheme, {\ncrypt}, which we carefully design in this work to work in synergy with $\SMac$. To the best of our knowledge, {\ncaudit} is the first scheme that efficiently supports auditing for NC storage systems. {\ncaudit} also provides protection against leakage of the outsourced data when the audit is done by a third party. Our evaluation results based on a real implementation in Java demonstrate that {\ncaudit} is significantly more efficient than the state-of-the-art schemes.


\balance


\appendix[Support for Data Dynamics]

{\ncaudit} supports data dynamics and does not require data block download (blockless) in all operations. The approach taken by {\ncaudit} is similar to \cite{Ateniese2008} but different from \cite{Erway2009} and \cite{QWang2009}:  {\ncaudit} fully supports block append and update operations, while relying on these two operations to further support insert and delete operations. Fully supporting all operations, as in \cite{Erway2009} and \cite{QWang2009}, come with a higher client and server computation as well as communication overhead. This is because additional data structures, such as a skip-list \cite{Erway2009} or a binary tree \cite{QWang2009}, must be maintained. We choose the simpler approach since data modification is typically of limited use for coded storage systems, as discussed in Section \ref{subsec:related-checking}.

{\flushleft \bf Block Append.}\quad Assume that the user wants to append a block, $\hat{\vct{b}}_{*}$, to the system. 
The coded blocks stored at the nodes are now a linear combination of the original source blocks, $\vct{b}_{1}, \cdots, \vct{b}_{m}$, and the new block $\vct{b}_{*}$. The encoded blocks stored at each node are updated based on the coding scheme used to attain the required level of reliability. 

\begin{figure}[tp]
\centering
\begin{tikzpicture}
\node[scale=0.7] {
\begin{tikzpicture}

\tikzstyle{post}=[
	->,
	shorten >=1pt,
	>=stealth',
	thick]

\tikzstyle{block} = [
	rectangle, 
      draw=black,
	fill=white, 
      thick,
	minimum width = 2.3cm]

\tikzstyle{storage} = [
	rectangle, 
      draw=black, 
	fill=blue!10,
      thick,
	minimum width = 2.7cm,
	minimum height = 2.3cm,
	rounded corners]

\node[storage]	(n1)	at (0, 0)				{};
\path (n1)+(0,+1.5) node (n1name) 	{Node 1};
\path (n1)+(+0,+0.72) node (b1) [block] 		{$\vct{b}_1$};
\node[block, below=0.5mm of b1]       (b2)   	{$\vct{b}_2$};

\node[storage]	(n2)	at (+3.5, 0)			{};
\path (n2)+(0,+1.5) node (n1name) 	{Node 2};
\path (n2)+(+0,+0.72) node (b3) [block] 		{$\vct{b}_3$};
\node[block, below=0.5mm of b3]       (b4)   	{$\vct{b}_4$};

\node[storage]	(n3)	at (+7, 0)			{};
\path (n3)+(0,+1.5) node (n1name) 	{Node 3};
\path (n3)+(+0,+0.72) node (b5) [block] 		{$\vct{b}_1 + \vct{b}_3$};
\node[block, below=0.5mm of b5]       (b6)  	 	{$\vct{b}_2 + \vct{b}_4$};

\node[storage]	(n4)	at (+10.5, 0)			{};
\path (n4)+(0,+1.5) node (n1name) 	{Node 4};
\path (n4)+(+0,+0.72) node (b7) [block] 		{$\vct{b}_2 + \vct{b}_3$};
\node[block, below=0.5mm of b7]       (b8)   	{$\vct{b}_1 + \vct{b}_2 + \vct{b}_4$};

\node[storage, below=1.5 of n1]		(n5)		{};
\path (n5)+(+0,+0.72) node (b9) [block] 	{$\vct{b}_1$};
\node[block, below=0.5mm of b9]       (b10)   	{$\vct{b}_2$};
\node[block, below=0.5mm of b10]       (empty)   	{ {\color{white} empty}};

\node[storage, below=1.5 of n2]		(n6)		{};
\path (n6)+(+0,+0.72) node (b11) [block] 	{$\vct{b}_3$};
\node[block, below=0.5mm of b11]       (b12)   	{$\vct{b}_4$};
\node[block, below=0.5mm of b12]       (b13)   	{\color{red} $\vct{b}_5$};

\node[storage, below=1.5 of n3]		(n7)		{};
\path (n7)+(+0,+0.72) node (b14) [block] 	{$\vct{b}_1 + \vct{b}_3$};
\node[block, below=0.5mm of b14]       (b15)   	{$\vct{b}_2 + \vct{b}_4$};
\node[block, below=0.5mm of b15]       (b16)   	{\color{red} $\vct{b}_5$};

\node[storage, below=1.5 of n4]		(n8)		{};
\path (n8)+(+0,+0.72) node (b17) [block] 	{\color{red} $\vct{b}_3$};
\node[block, below=0.5mm of b17]       (b18)   	{\color{red} $\vct{b}_1 + \vct{b}_4$};
\node[block, below=0.5mm of b18]       (b19)   	{\color{red} $\vct{b}_2 + \vct{b}_5$};
\draw[post] (b1.south) to (n8.north);
\draw[post] (b2.south) to (n8.north);

\end{tikzpicture}
};
\end{tikzpicture}
\caption{Appending a new block, $\vct{b}_5$, to an existing coded storage system (on top, as in Fig \ref{fig:repair}). The new system can still tolerate any two-node failure by leveraging an EVENODD code \cite{Blaum1994}. The user needs to upload $\vct{b}_5$ to nodes 2, 3, and 4, and instruct node 1 to sends to $\vct{b}_1$, $\vct{b}_2$, and their MAC tags to node 4.}
\label{fig:append}
\vspace*{-15pt}
\end{figure}
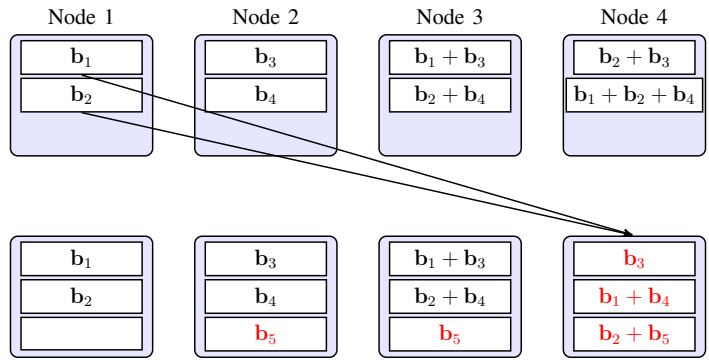

For instance, Fig. \ref{fig:append} shows how a new block, $\vct{b}_{5}$, could be added to an existing storage system (on top, as in Fig \ref{fig:repair}), where the new system can still tolerate any two-node failure by leveraging an EVENODD code \cite{Blaum1994}. Note that coded blocks at node 4 are completely changed. One way the user could achieve the new system is by instructing node 1 to send $\vct{b}_{1}$, $\vct{b}_{2}$, and their MAC tags to node 4 and also sending $\vct{b}_{5}$ to nodes 2, 3, and 4 him/herself. 

We focus our discussion on how security metadata can be maintained correctly and efficiently and assume that an appropriate update scheme for the data is in place, as shown in Fig. \ref{fig:append}. When an append is needed, the encoded $\vct{b}_{*}$ has the following form:
\[
\vct{b}_{*} = (\, \overbrace{\textrm{---}\vct{\hat{b}}_{*}\textrm{---}}^n, \overbrace{0, \cdots, 0}^m, 1)\,\in \mathbb{F}^{n+m+1}_q\,.
\]
To maintain the security metadata, the user first computes the tag $t_{\vct{b}_{*}}$ of $\vct{b}_{*}$ under $k_v$ using \mac~(now for vectors with size $n+m+1$) as follows:\\
\hspace*{2mm} -- $\vct{r}  \leftarrow (F_1 (k_v,  1), \cdots, F_1 (k_v,  n+m+1))$ .\\
\hspace*{2mm} -- $t_{\vct{b}_{*}} \leftarrow \vct{b}_{*} \cdot \vct{r} \in \mathbb{F}_q$ .\\
It then sends $t_{\vct{b}_{*}}$ to all storage nodes that have coded packets that involve $\vct{b}_{*}$ when sending  $\vct{b}_{*}$ to the nodes.

Note that when an append happens, the vector representation of a previous source block, $\vct{b}_i, i \in [1,m],$ is appended with a zero. However, its verification tag, computed using $\mac$, remains the same since $0 \times F_1 (k, n+m+1) = 0$. Consequently, for coded packets that do not involve $\vct{b}_{*}$, their tags remain the same, \ie, if $\vct{e} = \sum_{i=1}^{m} \alpha_i \, \vct{b}_i$, then its new tag equals is old tag: $t'_{\vct{e}} = t_{\vct{e}} = \sum_{i=1}^{m} \alpha_i \, t_{\vct{b}_i}$. For coded packets that involve $\vct{b}_{*}$, the storage node can compute their new tags using $t_{\vct{b}_{*}}$. Assume $\alpha_{*}$ of $\vct{b}_{*}$ is added to $\vct{e}$, then $t_{\vct{e}'} = t_{\vct{e}} + \alpha_{*} \, t_{\vct{b}_{*}}$. 

Afterwards, the user sends the new coding coefficients of the new coded blocks stored at the nodes to the TPA. Since the TPA carries out audits using this new set of coefficients, if the storage node does not update its data and tag correctly, it will not pass the subsequent audits. In particular, since the TPA computes $\aug{\vct{e}}$ in $\verifyproof$ locally, if the response block $\hat{\vct{e}}$ (before encryption) is not updated correctly, in the proof of Theorem \ref{thm:correctness}, $\vct{c} \neq \vct{e} + (\bar{\vct{m}}\,|\, 0, \cdots, 0)$. Thus, by the security guarantee of $\SMac$, $\verifyproof$ will fail w.h.p.

{\flushleft \bf Block Update.}\quad Assume the user wants to update the source block, $\vct{b}_j$, for some $j \in [1,m]$. Denote the new block after the update $\vct{b}'_j$. To update the data, it needs to send $\vct{b}'_j$ to nodes that store coded blocks involving $\vct{b}_j$. For example, to update $\vct{b}_3$ in Fig. \ref{fig:repair}, the user needs to send $\vct{b}'_3$ to the second, third, and fourth storage nodes so that they can update $\vct{b}_3$, $\vct{b}_1 + \vct{b}_3$, and $\vct{b}_2 + \vct{b}_3$, respectively.

To update the security metadata, the user first needs to learn the tag of $\vct{b}_j$, which can be done as follows. Assume $\vct{b}_j = \sum_{i=1}^m \alpha_i \, \vct{e}_i$, then $t_{\vct{b}_j} = \sum_{i=1}^m \alpha_i \, t_{\vct{e}_i}$. For $i \neq 0$, the user can download $t_{\vct{e}_i}$ from the appropriate storage nodes to compute $t_{\vct{b}_j}$. The user then computes the tag $t_{\vct{b}'_j}$ of $\vct{b}'_j$ under key $k_v$ using \mac. Finally, it sends the difference between $t_{\vct{b}'_j}$ and $t_{\vct{b}_j}$: $\delta_j = t_{\vct{b}'_j} - t_{\vct{b}_j}$, to the TPA.

Subsequently, whenever challenging a storage node and obtaining a response block which involves $\alpha_j \,\vct{b}_j$, the TPA runs $\verifyproof$ with the tag $t + \alpha_j \delta_j$ instead of $t$. To see why this is the case, let $\hat{\vct{e}} = \alpha_j \, \hat{\vct{b}}_j + \sum_{i=1, \cdots, M; i \neq j} \alpha_i \, \hat{\vct{b}}_i$ be the aggregated response block (before encryption). Its corresponding tag that is sent back with the proof of possession is $t = \alpha_j \, t_{\vct{b}_j} + \sum_{i=1, \cdots, M; i \neq j} \alpha_i \, t_{\vct{b}_i}$. But since $\vct{b}_j$ is now updated, the correct tag must be $t' = \alpha_j \, t_{\vct{b}'_j} + \sum_{i=1, \cdots, M; i \neq j} \alpha_i \, t_{\vct{b}_i} = t + \alpha_j \delta_j$. Note that if $\hat{\vct{e}}$ is not updated correctly by the storage node then by the security guarantee of $\SMac$, w.h.p. $t'$ is not a valid tag for $\vct{e}$. Subsequent updates to this $j$-th block can be carried out similarly.

This approach requires the TPA to store one field symbol $\delta_j$ for every updated source block $\vct{b}_j$, which is $O(m)$. This space overhead is negligible and could be constant in practice as discussed in Section \ref{audit-subsec:storageOverhead}. Finally, we assume that the storage nodes send back correct tags. If one wants to consider a stronger threat model where the storage nodes may send back corrupted tags, then there are two possible solutions: (i) modifying the auditing scheme to require the user to store the source tags, $t_{\vct{b}_j}$; in this case, the additional client storage overhead is $O(m)$ (still negligible); or (ii) a traditional MAC scheme computed on the coding coefficient, $\aug{\vct{e}_i}$, and verification tag, $t_{\vct{e}_i}$, can be used to protect the integrity of the tag.

{\flushleft \bf Block Insert.}\quad 
Similar to \cite{Ateniese2008}, a block insert is implemented with a block append and a mapping. In particular, the block is first appended to the system using {\em Block Append} above. Then the user needs to keep a mapping of the index of the appended block to its appropriate position. 

{\flushleft \bf Block Delete.}\quad We assume that the number of blocks to be deleted is small relatively to the file size. If a large portion of the file is to be deleted then it is best to rerun the {\em Setup} phase of {\ncaudit}. Similar to \cite{Ateniese2008}, we consider deletion of a block as changing it to a special block. Thus, deleting a block can be done as in the {\em Block Update} case.


\end{document}